\documentstyle[12pt]{article}
\begin{document}
\title{Asymptotically  Non-Static Kerr-deSitter Spacetime With No Event Horizon}
\author{Amir H. Abbassi*  and Sh. Khosravi\\
{\small Department of Physics, School of Sciences,}\\
{\small Tarbiat Modarres University,P.O. Box 14155-4838,Tehran , Iran.}\\
{\small * E-mail:ahabbasi@modares.ac.ir}}
\date{Dec. 2001}
\maketitle
\begin{abstract}
It has been shown in the presence of the cosmological constant any
solution of the Einstein field equations that asymptotically approaches
to the static desitter metric does not correspond to an observer of a
comoving frame but it should approach to the deSitter metric in the form
of Robertson - Walker metric (gr-qc/9812092) . For the case of
Schwarzschild - deSitter its proper form had been derived (gr-qc/9902009) .
Here we are going to present our derivations for a proper form of the
Kerr-deSitter metric. This has been done by considering the stationary
axially-symmetric spacetime in which motion of particle is integrable. That is the Hamilton-Jacobi and Klein-Gordon equations are separable. As in the Schwarzschild - deSitter case it does not possess any event horizon without imposing any additional conditions .Its intrinsic singularity and surfaces of infinite redshifts remains the same as common Kerr solutions.\\

\end{abstract}
\newpage
\section{Introduction}
A Positive cosmological constant can be inferred from the observations of high-redshift type Ia supernovae[1-6]. Previously we had shown that the static - deSitter metric is not acceptable according to the test of comoving coordinate frame by low magnitude - redshift relations[7]. For the Schwarzschild deSitter case the consistent form is [8] :
\begin{eqnarray}  
 ds^2&=&\frac{\Delta_{\rho}}{\rho^2}dt^2-\exp(2\sqrt{\frac{\Lambda}{3}}t)[\frac{\rho^{2}dr^2}{\Delta_{\rho}}-r^2(d\theta^2+\sin^{2}\theta\, d\phi^2)]\nonumber \\
\rho&\equiv&\exp(\sqrt{\frac{\Lambda}{3}}t)r\nonumber \\
\Delta_{\rho}&\equiv&\frac{1}{2}\left[(\rho^2-2M\rho-\frac{\Lambda}{3}\rho^4) + \sqrt{(\rho^2-2M\rho-\frac{\Lambda}{3}\rho^4)^2+\frac{4\Lambda}{3}\rho^6}\,\right]
\end{eqnarray}
This is evidently free of any event horizon and surface of infinite redshift for $\rho>0$.In the present work we intend to perform similar calculations for the case of Kerr space.The generalized Kerr metric in the presence of $\Lambda$ which asymptotically approaches to the static deSitter metric had been found by Carter[9,10]:

\begin{eqnarray}
ds^2 &=& (r^2+a^2\cos^{2}\theta)\left[\frac{dr^2}{\Delta_{r}}+\frac{d\theta^2}{1+\frac{\Lambda a^2}{3}\cos^{2}\theta}\right]+\nonumber \\
 & & \frac{\sin^{2}\theta(1+\frac{\Lambda a^2}{3}\cos^{2}\theta)}{(r^2+a^2\cos^{2}\theta)}\left[\frac{adt-(r^2+a^2)d\phi}{1+\frac{\Lambda a^2}{3}}\right]^2\nonumber \\
 & &-\frac{\Delta_{r}}{(r^2+a^2\cos^{2}\theta)}\left[\frac{dt-a\sin^{2}\theta d\phi}{1+\frac{\Lambda a^2}{3}}\right]^2\nonumber \\
\Delta_{r}&=&-\frac{\Lambda}{3}r^{2}(r^2+a^2)+r^2-2Mr+a^2 
\end{eqnarray}
We search for the event horizon by looking for the hypersurfaces where $r=constant$ becomes null, that is where $g^{11}$ vanishes.
The surfaces of infinite redshift in Kerr solution are given by the condition of vanishing of the coefficient $g_{00}$. Calculations of the Riemann invariant$R^{abcd}R_{abcd}$ for the Kerr metric , which can be obtained from (2) by substittuting $\Lambda=0$, reveals that it has one intrinsic singularity and it occurs when $r^2+\cos^{2}\theta=0$ . This leads to $r=\cos\theta=0$ which occurs when [11]
\begin{equation}
x^2+y^2=a^2 \,\, , \,\, z=0 .
\end{equation}
The surfaces of infinite redshift of Kerr metric are at
\begin{equation}
r=r_{s_{\pm}}=m\pm(m^2-a^2\cos^{2}\theta)^{\frac{1}{2}}
\end{equation}
and it has two null event horizons if $a^2<m^2$ ,
\begin{equation}
r=r_{\pm}=m\pm(m^2-a^2)^{\frac{1}{2}}
\end{equation}
The generalized Kerr metric (2) has an intrinsic singularity as (3) ,and two surfaces of infinite redshift and null event horizon very close to (4) and (5) respectively
\begin{equation}
r=r_{s_{\pm}}=m\pm[m^2-a^2\cos^{2}\theta(1+\frac{\Lambda a^2}{3}\sin^{2}\theta)]^{\frac{1}{2}}
\end{equation}
\begin{equation}
r=r_{\pm}\approx m\pm[m^2-a^2(1-\frac{\Lambda a^2}{3})]^{\frac{1}{2}}
\end{equation}
There are also an event horizon and surface of infinite redshift at very far distance, $r$ close to $\sqrt{\Lambda/3}$.

Derivation of the Kerr metric and the generalized Kerr solution is based on a method studying the separability of D'alambertian associated to the wave equation and integrability of its geodesic equations[9]. This separability depends not only on the geometry but on the particular choice of coordinates $x^a(a=0,1,2,3)$ and in terms of chosen coordinates this depends on the form of covariant tensor $g_{ab}$ defined by  
\begin{equation}
ds^2=g_{ab}dx^{a}dx^b,
\end{equation}
but in a less direct manner than on the form of the contravariant metric tensor $g^{ab}$ defined by the inverse co-form 
\begin{equation}
\left(\frac{\partial}{\partial s}\right)^2=g^{ab}\frac{\partial}{\partial x^a}\frac{\partial}{\partial x^b}.
\end{equation}
The Klein-Gordon equation can be expressed in terms of the components of contravariant metric tensor and the determinant
\begin{equation}
g=\det(g_{ab})={\det(g^{ab})}^{-1}
\end{equation}
The Klein-Gordon equation can be expressed in the form
\begin{equation}
\Psi^{-1}\frac{\partial}{\partial x^a}\sqrt{-g}\, g^{ab}\frac{\partial}{\partial x^b}\Psi-m^2\sqrt{-g}=0
\end{equation}
Standard separability takes place if substitution of wave function by the product $\Psi=\prod_{i}\psi_{i}$ of single-variable functions $\psi_{i}(i=0,\ldots ,3)$ of $x^{i}$ , causes the left-hand side of (11) to split up into four independent single-variable ordinary differential equations.We are going to use this method for deriving our new results.
\section{Derivations}
For the generalized Kerr metric one starts from the co-form
\begin{eqnarray}
\left(\frac{\partial}{\partial s}\right)^2&=&\frac{1}{Z}\left\{\Delta_{\mu}\left(\frac{\partial}{\partial\mu}\right)^2+\frac{1}{\Delta_{\mu}}\left[Z_{\mu}\frac{\partial}{\partial t}+Q_{\mu}\frac{\partial}{\partial\phi}\right]^2\right\}+ \nonumber \\
& &\frac{1}{Z}\left\{\Delta_{r}\left(\frac{\partial}{\partial r}\right)^2-\frac{1}{\Delta_{r}}\left[Z_{r}\frac{\partial}{\partial t}+Q_{r}\frac{\partial}{\partial\phi}\right]^2\right\}
\end{eqnarray}
where $\Delta_{\mu},Z_{\mu},Q_{\mu}$ are functions of $\mu=\cos\theta$ and $\Delta_{r},Z_{r}, Q_{r}$ are functions of $r$ only, and the form of conformal factor $Z$ remains to be determined. 
Now we are looking for a Kerr solution that asymptotically approaches to (1). It is more convenient to work in the coordinate system where $\rho=R(t)r$ is the radial coordinate. Then (1) becomes
\begin{eqnarray}
ds^2&=&\frac{1}{\rho^2}\left\{\left(\Delta_{\rho}-\frac{\frac{\Lambda\rho^6}{3}}{\Delta_{\rho}}\right)dt^2+2\sqrt{\frac{\Lambda}{3}}\rho^3 d\rho dt\right\}-\rho^2 \left\{\frac{d\rho^2}{\Delta_{\rho}}+d\theta^2+\sin^{2}\theta d\phi^2\right\} \nonumber \\
\Delta_{\rho}&=&\frac{1}{2}\left[\left(\rho^2-2M\rho-\frac{\Lambda}{3}\rho^4\right)+ \sqrt{(\rho^2-2M\rho-\frac{\Lambda}{3}\rho^4)^2+\frac{4\Lambda}{3}\rho^6}\,\right]
\end{eqnarray}
The determinant of $g_{ab}$ for (13) is equal to $-\rho^4$ that leads to $\sqrt{-g}=\rho^2$ and its corresponding form for obtaining the wave equation becomes
\begin{eqnarray}
\left(\frac{\partial}{\partial s}\right)^2&=&\frac{1}{\rho^2}\left\{(1-\mu^2)\left(\frac{\partial}{\partial \mu}\right)^2+\frac{1}{1-\mu^2}\left(\frac{\partial}{\partial\phi}\right)^2-2\sqrt{\frac{\Lambda}{3}}\rho^3\Delta_{\rho}\frac{\partial^2}{\partial t\partial \rho}+\right . \nonumber \\
& &\left .\left(\Delta_{\rho}-\frac{\frac{\Lambda\rho^6}{3}}{\Delta_{\rho}}\right)\left(\frac{\partial}{\partial\rho}\right)^2-\frac{\rho^4}{\Delta_{\rho}}\left(\frac{\partial}{\partial t}\right)^2\right\}.
\end{eqnarray}
where $\Delta_{\rho}$ is the same as given by (13).
Being asymptotically consistent with (13) and (14), instead of (12) we start with the following form
\begin{eqnarray}
\left(\frac{\partial}{\partial s}\right)^2&=&\frac{1}{Z}\left\{\Delta_{\mu}\left(\frac{\partial}{\partial\mu}\right)^2+\frac{1}{\Delta_{\mu}}\left[Z_{\mu}\frac{\partial}{\partial t}+Q_{\mu}\frac{\partial}{\partial\phi}\right]^2\right\}+ \nonumber \\
& &\frac{1}{Z}\left\{\Delta_{\rho}\left(\frac{\partial}{\partial \rho}\right)^2-\frac{1}{\Delta_{\rho}}\left[P\frac{\partial}{\partial\rho}+Z_{\rho}\frac{\partial}{\partial t}+Q_{\rho}\frac{\partial}{\partial\phi}\right]^2\right\}
\end{eqnarray}
where $P=\sqrt{\frac{\Lambda}{3}}\rho^3$ and the others are the same as (12). According to (15) the contravariant components of the metric are
\begin{eqnarray}
g^{\mu\mu}&=&\frac{\Delta_{\mu}}{Z}\,,\,g^{\phi\phi}=\frac{\frac{{Q_\mu}^2}{\Delta_\mu}-\frac{{Q_\rho}^2}{\Delta_\rho}}{Z}\,,\,g^{\rho\rho}=\frac{\Delta_\rho-\frac{P^2}{\Delta_\rho}}{Z}\,,\,g^{tt}=\frac{\frac{{Z_\mu}^2}{\Delta_\mu}-\frac{{Z_\rho}^2}{\Delta_\rho}}{Z} \nonumber \\
g^{t\phi}&=&\frac{\frac{Q_{\mu}Z_{\mu}}{\Delta_\mu}-\frac{Q_{\rho}Z_{\rho}}{\Delta_\rho}}{Z}\,,\,g^{t\rho}=-\frac{PZ_{\rho}}{\Delta_{\rho}Z}\,,\,g^{\rho\phi}=-\frac{PQ_{\rho}}{\Delta_{\rho}Z}
\end{eqnarray}
The determinant of $g_{ab}$ corresponding to (16) is
\begin{equation}
\det(g^{ab})=-\frac{(Q_{\mu}Z_{\rho}-Q_{\rho}Z_{\mu})^2}{Z^4}
\end{equation}
Since $\left(\det(g^{ab})\right)^{-1}=\det(g_{ab})=g$ ,(17) gives
\begin{equation}
\sqrt{-g}=\frac{Z^2}{|Z_{\rho}Q_{\mu}-Z_{\mu}Q_{\rho}|}
\end{equation}
In order to achieve the separability of the Klein-Gordon equation, we must impose a restriction on $Z$ so that $Z^{-1}\sqrt{-g}$ to be unity. By considering (18) we lead to choose
\begin{equation}
Z=Z_{\rho}Q_{\mu}-Z_{\mu}Q_{\rho}
\end{equation}
To complete the separability, the conformal factor $Z$ must split up into the sum of two parts each depending only on one variable. This requirement will be satisfied if and only if 
\begin{equation}
\frac{dZ_\rho}{d\rho}\frac{dQ_\mu}{d\mu}-\frac{dZ_\mu}{d\mu}\frac{dQ_\rho}{d\rho}=0
\end{equation}
Without loss of algebraic generality we may satisfy (20) by taking Q's to be constant. Thus replacing $Q_\rho$ and $Q_\mu$ by constants $C_\rho$ and $C_\mu$ respectively, we come to the basic separable canonical form for (15)
\begin{eqnarray}
\left(\frac{\partial}{\partial s}\right)^2&=&\frac{1}{C_{\mu}Z_{\rho}-C_{\rho}Z_{\mu}}\left\{\Delta_{\mu}\left(\frac{\partial}{\partial\mu}\right)^2+\frac{1}{\Delta_{\mu}}\left[Z_{\mu}\frac{\partial}{\partial t}+C_{\mu}\frac{\partial}{\partial\phi}\right]^2\right\}+ \nonumber \\
& &\frac{1}{C_{\mu}Z_{\rho}-C_{\rho}Z_{\mu}}\left\{\Delta_{\rho}\left(\frac{\partial}{\partial \rho}\right)^2-\frac{1}{\Delta_{\rho}}\left[P\frac{\partial}{\partial\rho}+Z_{\rho}\frac{\partial}{\partial t}+C_{\rho}\frac{\partial}{\partial\phi}\right]^2\right\}\nonumber \\
& &
\end{eqnarray}
The covariant metric corresponding to (21) is 
\begin{eqnarray}
ds^2&=&(C_{\mu}Z_{\rho}-C_{\rho}Z_{\mu})\left\{\frac{d\mu^2}{\Delta_\mu}+\frac{\left[d\rho-\frac{P(C_{\mu}dt-Z_{\mu}d\phi)}{C_{\mu}Z_{\rho}-C_{\rho}Z_{\mu}}\right]^2}{\Delta_\rho}\right\}+ \nonumber \\
& &\frac{1}{C_{\mu}Z_{\rho}-C_{\rho}Z_{\mu}}\left\{\Delta_{\mu}(C_{\rho}dt-Z_{\rho}d\phi)^2-\Delta_{\rho}(C_{\mu}dt-Z_{\mu}d\phi)^2\right\}
\end{eqnarray}
We may derive the solution corresponding to the cannonical form (22) by working in terms of the natural tetrad as follows
\begin{eqnarray}
\omega^{(1)}&=&\left(\frac{C_{\mu}Z_{\rho}-C_{\rho}Z_{\mu}}{\Delta_\rho}\right)^{\frac{1}{2}}\left[d\rho-\frac{P(C_{\mu}dt-Z_{\mu}d\phi)}{C_{\mu}Z_{\rho}-C_{\rho}Z_{\mu}}\right] \\
\omega^{(2)}&=&\left(\frac{C_{\mu}Z_{\rho}-C_{\rho}Z_{\mu}}{\Delta_\mu}\right)^{\frac{1}{2}}d\mu \\
\omega^{(3)}&=&\left(\frac{\Delta_\mu}{C_{\mu}Z_{\rho}-C_{\rho}Z_{\mu}}\right)^{\frac{1}{2}}(C_{\rho}dt-Z_{\rho}d\phi) \\
\omega^{(0)}&=&\left(\frac{\Delta_\rho}{C_{\mu}Z_{\rho}-C_{\rho}Z_{\mu}}\right)^{\frac{1}{2}}(C_{\mu}dt-Z_{\mu}d\phi) 
\end{eqnarray}
The four unknown functions $Z_\rho,Z_\mu,\Delta_\rho,\Delta_\mu$ should be determined in terms of $C_\mu,C_\rho,$ and $P=\sqrt{\frac{\Lambda}{3}}\rho^3$.

From (23)-(26) we have
\begin{eqnarray}
d\rho&=&\sqrt{\frac{\Delta_\rho}{C_{\mu}Z_{\rho}-C_{\rho}Z_{\mu}}}\omega^{(1)}+\frac{P\omega^{(0)}}{\sqrt{\Delta_\rho(C_{\mu}Z_{\rho}-C_{\rho}Z_{\mu})}} \\
d\mu&=&\sqrt{\frac{\Delta_\mu}{C_{\mu}Z_{\rho}-C_{\rho}Z_{\mu}}}\omega^{(2)} \\
d\phi&=&\frac{C_{\rho}\omega^{(0)}}{\sqrt{\Delta_\rho(C_{\mu}Z_{\rho}-C_{\rho}Z_{\mu})}}-\frac{C_{\mu}\omega^{(3)}}{\sqrt{\Delta_\mu(C_{\mu}Z_{\rho}-C_{\rho}Z_{\mu})}} \\
dt&=&\frac{Z_{\rho}\omega^{(0)}}{\sqrt{\Delta_\rho(C_{\mu}Z_{\rho}-C_{\rho}Z_{\mu})}}-\frac{Z_{\mu}\omega^{(3)}}{\sqrt{\Delta_\mu(C_{\mu}Z_{\rho}-C_{\rho}Z_{\mu})}} 
\end{eqnarray}
Derivatives of tetrads (or in other words the curl of $\omega^\mu$s), which are 1-forms, are themselves 2-forms $d\omega^\mu$ as follows
\begin{eqnarray}
d\omega^{(1)}&=&-\frac{C_{\rho}Z_{\mu}^{\prime}}{2Z}\sqrt{\frac{\Delta_\mu}{Z}}\omega^{(2)}\wedge\omega^{(1)}-\left[\frac{P^{\prime}}{\sqrt{Z\Delta_\rho}}-\frac{PC_{\mu}Z_{\rho}^{\prime}}{2Z\sqrt{Z\Delta_\rho}}-\frac{P\Delta_{\rho}^{\prime}}{2\Delta_{\rho}\sqrt{Z\Delta_\rho}}\right]\omega^{(1)}\wedge\omega^{(0)} \nonumber \\ 
& &-\frac{PC_{\mu}Z_{\mu}^{\prime}}{Z\sqrt{Z\Delta_\rho}}\omega^{(2)}\wedge\omega^{(3)} \\
d\omega^{(2)}&=&\frac{C_{\mu}Z_{\rho}^{\prime}}{2Z}\sqrt{\frac{\Delta_\rho}{Z}}\omega^{(1)}\wedge\omega^{(2)}+\frac{PC_{\mu}Z_{\rho}^{\prime}}{2Z\sqrt{Z\Delta_\rho}}\omega^{(0)}\wedge\omega^{(2)} \\
d\omega^{(3)}&=&\left(\frac{\Delta_{\mu}^{\prime}}{2\sqrt{Z\Delta_\mu}}+\frac{C_{\rho}Z_{\mu}^{\prime}}{2Z}\sqrt{\frac{\Delta_\mu}{Z}}\right)\omega^{(2)}\wedge\omega^{(3)}+\frac{C_{\mu}Z_{\rho}^{\prime}}{2Z}\sqrt{\frac{\Delta_\rho}{Z}}\omega^{(1)}\wedge\omega^{(3)} \nonumber \\
& &+\frac{PC_{\mu}Z_{\rho}^{\prime}}{2Z\sqrt{Z\Delta_\rho}}\omega^{(0)}\wedge\omega^{(3)}-\frac{C_{\rho}Z_{\rho}^{\prime}}{Z}\sqrt{\frac{\Delta_\mu}{Z}}\omega^{(1)}\wedge\omega^{(0)} \\
d\omega^{(0)}&=&\left(\frac{\Delta_{\rho}^{\prime}}{2\sqrt{Z\Delta_\rho}}-\frac{C_{\mu}Z_{\rho}^{\prime}}{2Z}\sqrt{\frac{\Delta_\rho}{Z}}\right)\omega^{(1)}\wedge\omega^{(0)}-\frac{C_{\rho}Z_{\mu}^{\prime}}{2Z}\sqrt{\frac{\Delta_\mu}{Z}}\omega^{(2)}\wedge\omega^{(0)}+ \nonumber \\
& &\frac{C_{\mu}Z_{\mu}^{\prime}}{Z}\sqrt{\frac{\Delta_\rho}{Z}}\omega^{(2)}\wedge\omega^{(3)}
\end{eqnarray}
where primes stand for derivatives with respect to the corresponding variable and $Z$ is given by (19).

By considering (31)-(34) and the first Cartan equation for zero torsion:
\begin{equation}
d\omega^\mu=-\omega^{(\mu)}_{\;\;\;(\nu)}\wedge\omega^{(\nu)}
\end{equation}
We have computed the 6 connection 1-forms $\omega^{(\mu)}_{\;\;\;(\nu)}$ where
\begin{equation}
dg_{\mu\nu}=\omega_{(\mu)(\nu)}+\omega_{(\nu)(\mu)}
\end{equation}
and in a basis that $g_{\mu\nu}$ is a constant (36) reduces to
\begin{equation}
\omega_{(\mu)(\nu)}=-\omega_{(\nu)(\mu)}
\end{equation}
The results are as follows
\begin{eqnarray}
\omega^{(1)}_{\;\;\;(0)}&=&\omega^{(0)}_{\;\;\;(1)}=\left(\frac{\Delta_{\rho}^{\prime}}{2\sqrt{Z\Delta_\rho}}-\frac{C_{\mu}Z_{\rho}^{\prime}}{2Z}\sqrt{\frac{\Delta_\rho}{Z}}\right)\omega^{(0)}+ \nonumber \\
& &\left(\frac{P^{\prime}}{\sqrt{Z\Delta_\rho}}-\frac{PC_{\mu}Z_{\rho}^{\prime}}{2Z\sqrt{Z\Delta_\rho}}-\frac{P\Delta_{\rho}^{\prime}}{2\Delta_{\rho}\sqrt{Z\Delta_\rho}}\right)\omega^{(1)}+\nonumber \\
& &\frac{C_{\rho}Z_{\rho}^{\prime}}{Z}\sqrt{\frac{\Delta_\mu}{Z}}\omega^{(3)} \\
\omega^{(1)}_{\;\;\;(2)}&=&-\omega^{(2)}_{\;\;\;(1)}=-\frac{C_{\rho}Z_{\mu}^{\prime}}{2Z}\sqrt{\frac{\Delta_\mu}{Z}}\omega^{(1)}-\frac{C_{\mu}Z_{\rho}^{\prime}}{2Z}\sqrt{\frac{\Delta_\rho}{Z}}\omega^{(2)}-\nonumber \\
& &\frac{PC_{\mu}Z_{\rho}^{\prime}}{2Z\sqrt{Z\Delta_\rho}}\omega^{(3)} \\
\omega^{(1)}_{\;\;\;(3)}&=&-\omega^{(3)}_{\;\;\;(1)}=\frac{C_{\rho}Z_{\rho}^{\prime}}{2Z}\sqrt{\frac{\Delta_\mu}{Z}}\omega^{(0)}+\frac{PC_{\mu}Z_{\mu}^{\prime}}{2Z\sqrt{Z\Delta_\rho}}\omega^{(2)}-\nonumber \\
& &\frac{C_{\mu}Z_{\rho}^{\prime}}{2Z}\sqrt{\frac{\Delta_\rho}{Z}}\omega^{(3)} \\
\omega^{(2)}_{\;\;\;(0)}&=&\omega^{(0)}_{\;\;\;(2)}=-\frac{C_{\rho}Z_{\mu}^{\prime}}{2Z}\sqrt{\frac{\Delta_\mu}{Z}}\omega^{(0)}+\frac{PC_{\mu}Z_{\rho}^{\prime}}{2Z\sqrt{Z\Delta_\rho}}\omega^{(2)}+\nonumber \\
& &\frac{C_{\mu}Z_{\mu}^{\prime}}{2Z}\sqrt{\frac{\Delta_\rho}{Z}}\omega^{(3)} \\
\omega^{(2)}_{\;\;\;(3)}&=&-\omega^{(3)}_{\;\;\;(2)}=\frac{C_{\mu}Z_{\mu}^{\prime}}{2Z}\sqrt{\frac{\Delta_\rho}{Z}}\omega^{(0)}+\frac{PC_{\mu}Z_{\mu}^{\prime}}{2Z\sqrt{Z\Delta_\rho}}\omega^{(1)}-\nonumber \\
& &\left(\frac{\Delta_{\mu}^{\prime}}{2\sqrt{Z\Delta_\mu}}+\frac{C_{\rho}Z_{\mu}^{\prime}}{2Z}\sqrt{\frac{\Delta_\mu}{Z}}\right)\omega^{(3)} \\
\omega^{(3)}_{\;\;\;(0)}&=&\omega^{(0)}_{\;\;\;(3)}=\frac{C_{\rho}Z_{\rho}^{\prime}}{2Z}\sqrt{\frac{\Delta_\mu}{Z}}\omega^{(1)}-\frac{C_{\mu}Z_{\mu}^{\prime}}{2Z}\sqrt{\frac{\Delta_\rho}{Z}}\omega^{(2)}+\frac{PC_{\mu}Z_{\rho}^{\prime}}{2Z\sqrt{Z\Delta_\rho}}\omega^{(3)} 
\end{eqnarray}
The curvature forms $\theta^{\mu}_{\;\;\nu}$ are defined as
\begin{equation}
\theta^{\mu}_{\;\;\nu}=d\omega^{(\mu)}_{\;\;\;(\nu)}+\omega^{(\mu)}_{\;\;\;(\alpha)}\wedge\omega^{(\alpha)}_{\;\;\;(\nu)}
\end{equation}
and according to the second Cartan equation we have
\begin{equation}
\theta^{\mu}_{\;\;\nu}=\frac{1}{2}R^{\mu}_{\nu\sigma\tau}\omega^{(\sigma)}\wedge\omega^{(\tau)}
\end{equation}
By computing $d\omega^{(\mu)}_{\;\;\;(\nu)}$ using (38)-(43)and inserting them in (44),$\theta^{\mu}_{\;\;\nu}$s are obtained as follows
\begin{eqnarray}
\theta^{1}_{\;\;0}&=&\theta^{0}_{\;\;1}=\left(\frac{\Delta_{\rho}^{\prime\prime}}{2Z}-\frac{\Delta_{\rho}^{\prime^{2}}}{4Z\Delta_\rho}-\frac{C_{\mu}Z_{\rho}^{\prime}\Delta_{\rho}^{\prime}}{2Z^2}-\frac{C_{\mu}Z_{\rho}^{\prime\prime}}{2Z^2}+\frac{3\left(C_{\mu}Z_{\rho}^{\prime}\right)^2}{4Z^3}\right)\omega^{(1)}\wedge\omega^{(0)}+ \nonumber \\
& &\left(\frac{C_{\rho}Z_{\mu}^{\prime}\Delta_{\rho}^{\prime}}{4Z^2}\sqrt{\frac{\Delta_\mu}{\Delta_\rho}}-\frac{3C_{\mu}Z_{\rho}^{\prime}C_{\rho}Z_{\mu}^{\prime}}{4Z^3}\sqrt{\Delta_{\rho}\Delta_{\mu}}\right)\omega^{(2)}\wedge\omega^{(0)}+ \nonumber \\
& &\left(\frac{\Delta_{\rho}^{\prime}}{2\sqrt{Z\Delta_\rho}}-\frac{C_{\mu}Z_{\rho}^{\prime}}{2Z}\sqrt{\frac{\Delta_\rho}{Z}}\right)\left[\left(\frac{\Delta_{\rho}^{\prime}}{2\sqrt{Z\Delta_\rho}}-\frac{C_{\mu}Z_{\rho}^{\prime}}{2Z}\sqrt{\frac{\Delta_\rho}{Z}}\right)\omega^{(1)}\wedge\omega^{(0)}\right.- \nonumber \\
& &\left.\frac{C_{\rho}Z_{\mu}^{\prime}}{2Z}\sqrt{\frac{\Delta_\mu}{Z}}\omega^{(2)}\wedge\omega^{(0)}+\frac{C_{\mu}Z_{\mu}^{\prime}}{Z}\sqrt{\frac{\Delta_\rho}{Z}}\omega^{(2)}\wedge\omega^{(3)}\right]+ \nonumber \\
& &\left(\frac{C_{\rho}Z_{\rho}^{\prime\prime}}{2Z^2}\sqrt{\Delta_{\rho}\Delta_{\mu}}-\frac{3C_{\rho}Z_{\rho}^{\prime}C_{\mu}Z_{\rho}^{\prime}}{4Z^3}\sqrt{\Delta_{\rho}\Delta_{\mu}}\right)\omega^{(1)}\wedge\omega^{(3)}+ \nonumber \\
& &P\left(\frac{C_{\rho}Z_{\rho}^{\prime\prime}}{2Z^2}\sqrt{\frac{\Delta_\mu}{\Delta_\rho}}-\frac{3C_{\rho}Z_{\rho}^{\prime}C_{\mu}Z_{\rho}^{\prime}}{4Z^3}\sqrt{\frac{\Delta_\mu}{\Delta_\rho}}\right)\omega^{(0)}\wedge\omega^{(3)}+ \nonumber \\
& &\left(\frac{C_{\rho}Z_{\rho}^{\prime}\Delta_{\mu}^{\prime}}{4Z^2}+\frac{3C_{\rho}Z_{\rho}^{\prime}C_{\rho}Z_{\mu}^{\prime}\Delta_{\mu}}{4Z^3}\right)\omega^{(2)}\wedge\omega^{(3)}+ \nonumber \\
& &\frac{C_{\rho}Z_{\rho}^{\prime}}{2Z}\sqrt{\frac{\Delta_\mu}{Z}}\left[\left(\frac{\Delta_{\mu}^{\prime}}{2\sqrt{Z\Delta_\mu}}+\frac{C_{\rho}Z_{\mu}^{\prime}}{2Z}\sqrt{\frac{\Delta_\mu}{Z}}\right)\omega^{(2)}\wedge\omega^{(3)}+\frac{C_{\mu}Z_{\rho}^{\prime}}{2Z}\sqrt{\frac{\Delta_\rho}{Z}}\omega^{(1)}\wedge\omega^{(3)}+\right. \nonumber \\
& &\left.\frac{C_{\mu}Z_{\rho}^{\prime}P}{2Z\sqrt{Z\Delta_\rho}}\omega^{(0)}\wedge\omega^{(3)}-\frac{C_{\rho}Z_{\rho}^{\prime}}{Z}\sqrt{\frac{\Delta_\rho}{Z}}\omega^{(1)}\wedge\omega^{(0)}\right]+ \nonumber \\
& &P\left[\frac{P^{\prime\prime}}{Z\Delta_\rho}-\frac{P^{\prime}\Delta_{\rho}^{\prime}}{\Delta_{\rho}^{2}Z}-\frac{P^{\prime}C_{\mu}Z_{\rho}^{\prime}}{Z^{2}\Delta_{\rho}}-\frac{PC_{\mu}Z_{\rho}^{\prime\prime}}{2Z^{2}\Delta_{\rho}}+\frac{3P(C_{\mu}Z_{\rho}^{\prime})^2}{4Z^{3}\Delta_{\rho}}+\right.\nonumber \\
& &\left.\frac{PC_{\mu}Z_{\rho}^{\prime}\Delta_{\rho}^{\prime}}{2Z^{2}\Delta_{\rho}^{2}}-\frac{P\Delta_{\rho}^{\prime\prime}}{2Z\Delta_{\rho}^{2}}+\frac{3P\Delta_{\rho}^{\prime^{2}}}{4\Delta_{\rho}^{3}Z}\right]\omega^{(0)}\wedge\omega^{(1)}+ \nonumber \\
& &\left(\frac{P^{\prime}C_{\rho}Z_{\mu}^{\prime}}{2Z^{2}}\sqrt{\frac{\Delta_{\mu}}{\Delta_{\rho}}}-\frac{3PC_{\mu}Z_{\rho}^{\prime}C_{\rho}Z_{\mu}^{\prime}}{4Z^3}\sqrt{\frac{\Delta_{\mu}}{\Delta_{\rho}}}-\frac{PC_{\rho}Z_{\mu}^{\prime}\Delta_{\rho}^{\prime}}{4\Delta_{\rho}Z^2}\sqrt{\frac{\Delta_{\mu}}{\Delta_{\rho}}}\right)\omega^{(2)}\wedge\omega^{(1)}+ \nonumber \\
& &\left(\frac{P^{\prime}}{\sqrt{Z\Delta_{\rho}}}-\frac{PC_{\mu}Z_{\rho}^{\prime}}{2Z\sqrt{Z\Delta_\rho}}-\frac{P\Delta_{\rho}^{\prime}}{2\Delta_{\rho}\sqrt{Z\Delta_\rho}}\right)\left[-\frac{C_{\rho}Z_{\mu}^{\prime}}{2Z}\sqrt{\frac{\Delta_\mu}{Z}}\omega^{(2)}\wedge\omega^{(1)}-\right. \nonumber \\
& &\left. \left(\frac{P^{\prime}}{\sqrt{Z\Delta_{\rho}}}-\frac{PC_{\mu}Z_{\rho}^{\prime}}{2Z\sqrt{Z\Delta_\rho}}-\frac{P\Delta_{\rho}^{\prime}}{2\Delta_{\rho}\sqrt{Z\Delta_\rho}}\right)\omega^{(1)}\wedge\omega^{(0)}-\frac{PC_{\mu}Z_{\mu}^{\prime}}{Z\sqrt{Z\Delta_\rho}}\omega^{(2)}\wedge\omega^{(3)}\right] +  \nonumber \\
& &\left(-\frac{C_{\rho}Z_{\mu}^{\prime}}{2Z}\sqrt{\frac{\Delta_\mu}{Z}}\omega^{(1)}-\frac{C_{\mu}Z_{\rho}^{\prime}}{2Z}\sqrt{\frac{\Delta_\rho}{Z}}\omega^{(2)}-\frac{PC_{\mu}Z_{\mu}^{\prime}}{2Z\sqrt{Z\Delta_\rho}}\omega^{(3)}\right)\wedge  \nonumber \\
& &\left(-\frac{C_{\rho}Z_{\mu}^{\prime}}{2Z}\sqrt{\frac{\Delta_\mu}{Z}}\omega^{(0)}+\frac{PC_{\mu}Z_{\rho}^{\prime}}{2Z\sqrt{Z\Delta_\rho}}\omega^{(2)}+\frac{C_{\mu}Z_{\mu}^{\prime}}{2Z}\sqrt{\frac{\Delta_\rho}{Z}}\omega^{(3)}\right)+ \nonumber \\
& &\left(\frac{C_{\rho}Z_{\rho}^{\prime}}{2Z}\sqrt{\frac{\Delta_\mu}{Z}}\omega^{(0)}+\frac{PC_{\mu}Z_{\mu}^{\prime}}{2Z\sqrt{Z\Delta_\rho}}\omega^{(2)}-\frac{C_{\mu}Z_{\rho}^{\prime}}{2Z}\sqrt{\frac{\Delta_\rho}{Z}}\omega^{(3)}\right)\wedge  \nonumber \\
& &\left(\frac{C_{\rho}Z_{\rho}^{\prime}}{2Z}\sqrt{\frac{\Delta_\mu}{Z}}\omega^{(1)}-\frac{C_{\mu}Z_{\mu}^{\prime}}{2Z}\sqrt{\frac{\Delta_\rho}{Z}}\omega^{(2)}+\frac{PC_{\mu}Z_{\rho}^{\prime}}{2Z\sqrt{Z\Delta_\rho}}\omega^{(3)}\right) \\
\theta^{2}_{\;\;0}&=&\theta^{0}_{\;\;2}=\left(-\frac{C_{\rho}Z_{\mu}^{\prime\prime}\Delta_{\mu}}{2Z^2}-\frac{C_{\rho}Z_{\mu}^{\prime}\Delta_{\mu}^{\prime}}{4Z^2}-\frac{3(C_{\rho}Z_{\mu}^{\prime})^2\Delta_{\mu}}{4Z^3}\right)\omega^{(2)}\wedge\omega^{(0)}+ \nonumber \\
& &\frac{3C_{\rho}Z_{\rho}^{\prime}C_{\mu}Z_{\mu}^{\prime}}{4Z^3}\sqrt{\Delta_{\rho}\Delta_{\mu}}\omega^{(1)}\wedge\omega^{(0)}-\nonumber \\
& &\frac{C_{\rho}Z_{\mu}^{\prime}}{2Z}\sqrt{\frac{\Delta_\mu}{Z}}\left[\left(\frac{\Delta_{\rho}^{\prime}}{2\sqrt{Z\Delta_\rho}}-\frac{C_{\mu}Z_{\rho}^{\prime}}{2Z}\sqrt{\frac{\Delta_\rho}{Z}}\right)\omega^{(1)}\wedge\omega^{(0)}\right.- \nonumber \\
& &\left.\frac{C_{\rho}Z_{\mu}^{\prime}}{2Z}\sqrt{\frac{\Delta_\mu}{Z}}\omega^{(2)}\wedge\omega^{(0)}+\frac{C_{\mu}Z_{\mu}^{\prime}}{Z}\sqrt{\frac{\Delta_\rho}{Z}}\omega^{(2)}\wedge\omega^{(3)}\right]+ \nonumber \\
& &\left(\frac{C_{\mu}Z_{\mu}^{\prime\prime}}{2Z^2}\sqrt{\Delta_{\rho}\Delta_{\mu}}+\frac{3C_{\mu}Z_{\mu}^{\prime}C_{\rho}Z_{\mu}^{\prime}}{4Z^3}\sqrt{\Delta_{\rho}\Delta_{\mu}}\right)\omega^{(2)}\wedge\omega^{(3)}+ \nonumber \\
& &\left(\frac{C_{\mu}Z_{\mu}^{\prime}\Delta_{\rho}^{\prime}}{4Z^2}-\frac{3C_{\mu}Z_{\rho}^{\prime}C_{\mu}Z_{\mu}^{\prime}\Delta_{\rho}}{4Z^3}\right)\omega^{(1)}\wedge\omega^{(3)}+\nonumber \\
& &P\left(\frac{C_{\mu}Z_{\mu}^{\prime}\Delta_{\rho}^{\prime}}{4Z^{2}\Delta_{\rho}}-\frac{3C_{\mu}Z_{\rho}^{\prime}C_{\mu}Z_{\mu}^{\prime}}{4Z^3}\right)\omega^{(0)}\wedge\omega^{(3)}+ \nonumber \\
& &\frac{C_{\mu}Z_{\mu}^{\prime}}{2Z}\sqrt{\frac{\Delta_\rho}{Z}}\left[\left(\frac{\Delta_{\mu}^{\prime}}{2\sqrt{Z\Delta_\mu}}+\frac{C_{\rho}Z_{\mu}^{\prime}}{2Z}\sqrt{\frac{\Delta_\mu}{Z}}\right)\omega^{(2)}\wedge\omega^{(3)}+\right. \nonumber \\
& &\left.\frac{C_{\mu}Z_{\rho}^{\prime}}{2Z}\sqrt{\frac{\Delta_\rho}{Z}}\omega^{(1)}\wedge\omega^{(3)}+
\frac{C_{\mu}Z_{\rho}^{\prime}P}{2Z\sqrt{Z\Delta_\rho}}\omega^{(0)}\wedge\omega^{(3)}-\frac{C_{\rho}Z_{\rho}^{\prime}}{Z}\sqrt{\frac{\Delta_\mu}{Z}}\omega^{(1)}\wedge\omega^{(0)}\right]+ \nonumber \\
& &\left[\frac{-3(C_{\mu}Z_{\rho}^{\prime})^{2}P}{4Z^3}+\frac{C_{\mu}Z_{\rho}^{\prime\prime}P}{2Z^2}+\frac{C_{\mu}Z_{\rho}^{\prime}P^{\prime}}{2Z^2}-\frac{C_{\mu}Z_{\rho}^{\prime}P\Delta_{\rho}^{\prime}}{4Z^{2}\Delta_{\rho}}\right]\omega^{(1)}\wedge\omega^{(2)}+ \nonumber \\
& &\frac{P}{\Delta_{\rho}}\left[\frac{-3(C_{\mu}Z_{\rho}^{\prime})^{2}P}{4Z^3}+\frac{C_{\mu}Z_{\rho}^{\prime\prime}P}{2Z^2}+\frac{C_{\mu}Z_{\rho}^{\prime}P^{\prime}}{2Z^2}-\frac{C_{\mu}Z_{\rho}^{\prime}P\Delta_{\rho}^{\prime}}{4Z^{2}\Delta_{\rho}}\right]\omega^{(0)}\wedge\omega^{(2)}+ \nonumber \\
& &\frac{C_{\mu}Z_{\rho}^{\prime}P}{2Z\sqrt{Z\Delta_{\rho}}}\left[\frac{C_{\mu}Z_{\rho}^{\prime}}{2Z}\sqrt{\frac{\Delta_{\rho}}{Z}}\omega^{(1)}\wedge\omega^{(2)}+\frac{C_{\mu}Z_{\rho}^{\prime}P}{2Z\sqrt{Z\Delta_{\rho}}}\omega^{(0)}\wedge\omega^{(2)}\right]+\nonumber \\
& &\left(\frac{C_{\rho}Z_{\mu}^{\prime}}{2Z}\sqrt{\frac{\Delta_{\mu}}{Z}}\omega^{(1)}+\frac{C_{\mu}Z_{\rho}^{\prime}}{2Z}\sqrt{\frac{\Delta_{\rho}}{Z}}\omega^{(2)}+\frac{C_{\mu}Z_{\mu}^{\prime}P}{2Z\sqrt{Z\Delta_{\rho}}}\omega^{(3)}\right)\wedge\nonumber \\
& &\left[\left(\frac{\Delta_{\rho}^{\prime}}{2\sqrt{Z\Delta_\rho}}-\frac{C_{\mu}Z_{\rho}^{\prime}}{2Z}\sqrt{\frac{\Delta_\rho}{Z}}\right)\omega^{(0)}+\right.\nonumber \\
& &\left.\left(\frac{P^{\prime}}{\sqrt{Z\Delta_{\rho}}}-\frac{PC_{\mu}Z_{\rho}^{\prime}}{2Z\sqrt{Z\Delta_{\rho}}}-\frac{P\Delta_{\rho}^{\prime}}{2\Delta_{\rho}\sqrt{Z\Delta_{\rho}}}\right)\omega^{(1)}+\frac{C_{\rho}Z_{\rho}^{\prime}}{2Z}\sqrt{\frac{\Delta_{\mu}}{Z}}\omega^{(3)}\right]+\nonumber \\
& &\left[\frac{C_{\mu}Z_{\mu}^{\prime}}{2Z}\sqrt{\frac{\Delta_{\rho}}{Z}}\omega^{(0)}+\frac{C_{\mu}Z_{\mu}^{\prime}P}{2Z\sqrt{Z\Delta_{\rho}}}\omega^{(1)}-\left(\frac{\Delta_{\mu}^{\prime}}{2\sqrt{Z\Delta_\mu}}+\frac{C_{\rho}Z_{\mu}^{\prime}}{2Z}\sqrt{\frac{\Delta_\mu}{Z}}\right)\omega^{(3)}\right]\wedge \nonumber \\
& &\left(\frac{C_{\rho}Z_{\rho}^{\prime}}{2Z}\sqrt{\frac{\Delta_{\mu}}{Z}}\omega^{(1)}-\frac{C_{\mu}Z_{\mu}^{\prime}}{2Z}\sqrt{\frac{\Delta_{\rho}}{Z}}\omega^{(2)}+\frac{C_{\mu}Z_{\rho}^{\prime}P}{2Z\sqrt{Z\Delta_{\rho}}}\omega^{(3)}\right) \\
\theta^{3}_{\;\;0}&=&\theta^{0}_{\;\;3}=\left(\frac{C_{\rho}Z_{\rho}^{\prime}}{4Z^2}\Delta_{\mu}^{\prime}+\frac{3C_{\rho}Z_{\rho}^{\prime}C_{\rho}Z_{\mu}^{\prime}}{4Z^3}\Delta_{\mu}\right)\omega^{(2)}\wedge\omega^{(1)}+\nonumber \\
& &P\sqrt{\frac{\Delta_{\mu}}{\Delta_{\rho}}}\left(\frac{C_{\rho}Z_{\rho}^{\prime\prime}}{2Z^2}-\frac{3C_{\rho}Z_{\rho}^{\prime}C_{\mu}Z_{\rho}^{\prime}}{4Z^3}\right)\omega^{(0)}\wedge\omega^{(1)}+\nonumber \\
& &\frac{C_{\rho}Z_{\rho}^{\prime}}{2Z}\sqrt{\frac{\Delta_{\mu}}{Z}}\left[-\frac{C_{\rho}Z_{\mu}^{\prime}}{2Z}\omega^{(2)}\wedge\omega^{(1)}-\right.\nonumber \\
& &\left.\left(\frac{P^\prime}{\sqrt{Z\Delta_\rho}}-\frac{PC_{\mu}Z_{\rho}^{\prime}}{2Z\sqrt{Z\Delta_\rho}}-\frac{P\Delta_{\rho}^{\prime}}{2\Delta_{\rho}\sqrt{Z\Delta_\rho}}\right)\omega^{(1)}\wedge\omega^{(0)}\right.-\nonumber \\
& &\left.\frac{PC_{\mu}Z_{\mu}^{\prime}}{Z\sqrt{Z\Delta_\rho}}\omega^{(2)}\wedge\omega^{(3)}\right]+\left(-\frac{C_{\mu}Z_{\mu}^{\prime}}{2Z^2}\Delta_{\rho}^{\prime}+\frac{3C_{\mu}Z_{\rho}^{\prime}C_{\mu}Z_{\mu}^{\prime}}{4Z^3}\Delta_{\rho}\right)\omega^{(1)}\wedge\omega^{(2)}+\nonumber \\
& &\frac{P}{\Delta_\rho}\left(-\frac{C_{\mu}Z_{\mu}^{\prime}}{2Z^2}\Delta_{\rho}^{\prime}+\frac{3C_{\mu}Z_{\rho}^{\prime}C_{\mu}Z_{\mu}^{\prime}}{4Z^3}\Delta_{\rho}\right)\omega^{(0)}\wedge\omega^{(2)}-\nonumber \\
& &\frac{C_{\mu}Z_{\mu}^{\prime}}{2Z}\sqrt{\frac{\Delta_{\rho}}{Z}}\left[+\frac{C_{\mu}Z_{\rho}^{\prime}}{2Z}\sqrt{\frac{\Delta_{\rho}}{Z}}\omega^{(1)}\wedge\omega^{(2)}+\frac{PC_{\mu}Z_{\rho}^{\prime}}{2Z\sqrt{Z\Delta_\rho}}\omega^{(0)}\wedge\omega^{(2)}\right]+\nonumber \\
& &\frac{P}{\Delta_\rho}\left(\frac{C_{\mu}Z_{\rho}^{\prime}P^{\prime}}{2Z^2}-\frac{C_{\mu}Z_{\rho}^{\prime}\Delta_{\rho}^{\prime}P}{4Z^2\Delta_{\rho}}-\frac{3(C_{\mu}Z_{\rho}^{\prime})^2P}{4Z^3}+\frac{C_{\mu}Z_{\rho}^{\prime\prime}P}{2Z^2}\right)\omega^{(0)}\wedge\omega^{(3)}+\nonumber \\
& &\frac{3C_{\mu}Z_{\rho}^{\prime}C_{\rho}Z_{\mu}^{\prime}P}{4Z^3}\sqrt{\frac{\Delta_{\mu}}{\Delta_{\rho}}}\omega^{(2)}\wedge\omega^{(3)}+\nonumber \\
& &\frac{PC_{\mu}Z_{\rho}^{\prime}}{2Z\sqrt{Z\Delta_\rho}}\left[\left(\frac{\Delta_{\mu}^{\prime}}{2\sqrt{Z\Delta_{\mu}}}+\frac{C_{\rho}Z_{\mu}^{\prime}}{2Z}\sqrt{\frac{\Delta_{\mu}}{Z}}\right)\omega^{(2)}\wedge\omega^{(3)}\right.+\nonumber \\
& &\left.\frac{C_{\mu}Z_{\rho}^{\prime}}{2Z}\sqrt{\frac{\Delta_{\rho}}{Z}}\omega^{(1)}\wedge\omega^{(3)}+\frac{PC_{\mu}Z_{\rho}^{\prime}}{2Z\sqrt{Z\Delta_\rho}}\omega^{(0)}\wedge\omega^{(3)}-\frac{C_{\rho}Z_{\rho}^{\prime}}{Z}\sqrt{\frac{\Delta_{\mu}}{Z}}\omega^{(1)}\wedge\omega^{(0)}\right]+\nonumber \\
& &\left(-\frac{C_{\rho}Z_{\rho}^{\prime}}{2Z}\sqrt{\frac{\Delta_{\mu}}{Z}}\omega^{(0)}-\frac{PC_{\mu}Z_{\mu}^{\prime}}{2Z\sqrt{Z\Delta_\rho}}\omega^{(2)}+\frac{C_{\mu}Z_{\rho}^{\prime}}{2Z}\sqrt{\frac{\Delta_{\rho}}{Z}}\omega^{(3)}\right)\wedge\nonumber \\
& &\left[\left(\frac{\Delta_{\rho}^{\prime}}{2\sqrt{Z\Delta_{\rho}}}-\frac{C_{\mu}Z_{\rho}^{\prime}}{2Z}\sqrt{\frac{\Delta_{\rho}}{Z}}\right)\omega^{(0)}+\right.\nonumber \\
& &\left.\left(\frac{P^\prime}{\sqrt{Z\Delta_\rho}}-\frac{PC_{\mu}Z_{\rho}^{\prime}}{2Z\sqrt{Z\Delta_\rho}}-\frac{P\Delta_{\rho}^{\prime}}{2\Delta_{\rho}\sqrt{Z\Delta_\rho}}\right)\omega^{(1)}\right.+\nonumber \\
& &\left.\frac{C_{\rho}Z_{\rho}^{\prime}}{2Z}\sqrt{\frac{\Delta_{\mu}}{Z}}\omega^{(3)}\right]+\left[-\frac{C_{\mu}Z_{\mu}^{\prime}}{2Z}\sqrt{\frac{\Delta_{\rho}}{Z}}\omega^{(0)}-\frac{PC_{\mu}Z_{\mu}^{\prime}}{2Z\sqrt{Z\Delta_\rho}}\omega^{(1)}+\right.\nonumber \\
& &\left.\left(\frac{\Delta_{\mu}^{\prime}}{2\sqrt{Z\Delta_{\mu}}}-\frac{C_{\rho}Z_{\mu}^{\prime}}{2Z}\sqrt{\frac{\Delta_{\mu}}{Z}}\right)\omega^{(3)}\right]\wedge \nonumber \\
& &\left[-\frac{C_{\rho}Z_{\mu}^{\prime}}{2Z}\sqrt{\frac{\Delta_{\rho}}{Z}}\omega^{(0)}+\frac{PC_{\mu}Z_{\rho}^{\prime}}{2Z\sqrt{Z\Delta_\rho}}\omega^{(2)}+\frac{C_{\mu}Z_{\mu}^{\prime}}{2Z}\sqrt{\frac{\Delta_{\rho}}{Z}}\omega^{(3)}\right] \\
\theta^{2}_{\;\;1}&=&-\theta^{1}_{\;\;2}=\left(\frac{C_{\rho}Z_{\mu}^{\prime\prime}\Delta_{\mu}}{2Z^2}+\frac{C_{\rho}Z_{\mu}^{\prime}\Delta_{\mu}^{\prime}}{4Z^2}+\frac{3(C_{\rho}Z_{\mu}^{\prime})^2\Delta_{\mu}}{4Z^3}\right)\omega^{(2)}\wedge\omega^{(1)}- \nonumber \\
& &\frac{3C_{\rho}Z_{\mu}^{\prime}C_{\mu}Z_{\rho}^{\prime}P}{4Z^3}\sqrt{\frac{\Delta_\mu}{\Delta_\rho}}\omega^{(0)}\wedge\omega^{(1)}+\frac{C_{\rho}Z_{\mu}^{\prime}}{2Z}\sqrt{\frac{\Delta_{\mu}}{Z}}\left[-\frac{C_{\rho}Z_{\mu}^{\prime}}{2Z}\sqrt{\frac{\Delta_{\mu}}{Z}}\omega^{(2)}\wedge\omega^{(1)}-\right. \nonumber \\
& &\left.\left(\frac{P^\prime}{Z\Delta_\rho}-\frac{PC_{\mu}Z_{\rho}^{\prime}}{2Z\sqrt{Z\Delta_\rho}}-\frac{P\Delta_{\rho}^{\prime}}{2\Delta_{\rho}\sqrt{Z\Delta_\rho}}\right)\omega^{(1)}\wedge\omega^{(0)}-\frac{PC_{\mu}Z_{\mu}^{\prime}}{Z\sqrt{Z\Delta_\rho}}\omega^{(2)}\wedge\omega^{(3)}\right]+ \nonumber \\
& &\left(\frac{C_{\mu}Z_{\rho}^{\prime\prime}\Delta_{\rho}}{2Z^2}+\frac{C_{\mu}Z_{\rho}^{\prime}\Delta_{\rho}^{\prime}}{4Z^2}-\frac{3(C_{\mu}Z_{\rho}^{\prime})^2\Delta_{\rho}}{4Z^3}\right)\omega^{(1)}\wedge\omega^{(2)}+ \nonumber \\
& &\frac{P}{\Delta_\rho}\left(\frac{C_{\mu}Z_{\rho}^{\prime\prime}\Delta_{\rho}}{2Z^2}+\frac{C_{\mu}Z_{\rho}^{\prime}\Delta_{\rho}^{\prime}}{4Z^2}-\frac{3(C_{\mu}Z_{\rho}^{\prime})^2\Delta_{\rho}}{4Z^3}\right)\omega^{(0)}\wedge\omega^{(2)}+ \nonumber \\
& &\frac{C_{\mu}Z_{\rho}^{\prime}}{2Z}\sqrt{\frac{\Delta_{\rho}}{z}}\left[\frac{C_{\mu}Z_{\rho}^{\prime}}{2Z}\sqrt{\frac{\Delta_{\rho}}{Z}}\omega^{(1)}\wedge\omega^{(2)}+\frac{PC_{\mu}Z_{\rho}^{\prime}}{2Z\sqrt{Z\Delta_\rho}}\omega^{(0)}\wedge\omega^{(2)}\right]+ \nonumber \\
& &\left(\frac{PC_{\mu}Z_{\mu}^{\prime\prime}}{2Z^2}\sqrt{\frac{\Delta_\mu}{\Delta_\rho}}+\frac{3C_{\rho}Z_{\mu}^{\prime}C_{\mu}Z_{\mu}^{\prime}P}{4Z^3}\sqrt{\frac{\Delta_\mu}{\Delta_\rho}}\right)\omega^{(2)}\wedge\omega^{(3)}+ \nonumber \\
& &\left(\frac{P^{\prime}C_{\mu}Z_{\mu}^{\prime}}{2Z^2}-\frac{PC_{\mu}Z_{\mu}^{\prime}\Delta_{\rho}^{\prime}}{4Z^{2}\Delta_{\rho}}-\frac{3C_{\mu}Z_{\mu}^{\prime}C_{\mu}Z_{\rho}^{\prime}P}{4Z^3}\right)\omega^{(1)}\wedge\omega^{(3)}+ \nonumber \\
& &\frac{P}{\Delta_\rho}\left(\frac{P^{\prime}C_{\mu}Z_{\mu}^{\prime}}{2Z^2}-\frac{PC_{\mu}Z_{\mu}^{\prime}\Delta_{\mu}^{\prime}}{4Z^{2}\Delta_{\rho}}-\frac{3C_{\mu}Z_{\mu}^{\prime}C_{\mu}Z_{\rho}^{\prime}P}{4Z^3}\right)\omega^{(0)}\wedge\omega^{(3)}+ \nonumber \\
& &\frac{PC_{\mu}Z_{\mu}^{\prime}}{Z\sqrt{2Z\Delta_\rho}}\left[\left(\frac{\Delta_{\mu}^{\prime}}{2\sqrt{Z\Delta_\mu}}+\frac{C_{\rho}Z_{\mu}^{\prime}}{2Z}\sqrt{\frac{\Delta_{\mu}}{z}}\right)\omega^{(2)}\wedge\omega^{(3)}+\frac{C_{\mu}Z_{\rho}^{\prime}}{2Z}\sqrt{\frac{\Delta_{\rho}}{Z}}\omega^{(1)}\wedge\omega^{(3)}+\right. \nonumber \\
& &\left.\frac{PC_{\mu}Z_{\rho}^{\prime}}{2Z\sqrt{Z\Delta_\rho}}\omega^{(0)}\wedge\omega^{(3)}-\frac{C_{\rho}Z_{\rho}^{\prime}}{Z}\sqrt{\frac{\Delta_{\mu}}{z}}\omega^{(1)}\wedge\omega^{(0)}\right]+ \nonumber \\
& &\left(-\frac{C_{\rho}Z_{\mu}^{\prime}}{2Z}\sqrt{\frac{\Delta_{\mu}}{z}}\omega^{(0)}+\frac{PC_{\mu}Z_{\rho}^{\prime}}{2Z\sqrt{Z\Delta_\rho}}\omega^{(2)}+\frac{C_{\mu}Z_{\mu}^{\prime}}{2Z}\sqrt{\frac{\Delta_{\rho}}{Z}}\omega^{(3)}\right)\wedge\nonumber\ \\
& &\left[\left(\frac{\Delta_{\rho}^{\prime}}{2\sqrt{Z\Delta_\rho}}-\frac{C_{\mu}Z_{\rho}^{\prime}}{2Z}\sqrt{\frac{\Delta_{\rho}}{Z}}\right)\omega^{(0)}+\right.\nonumber \\
& &\left.\left(\frac{P^{\prime}}{\sqrt{Z\Delta_\rho}}-\frac{PC_{\mu}Z_{\rho}^{\prime}}{2Z\sqrt{Z\Delta_\rho}}-\frac{P\Delta_{\rho}^{\prime}}{2\Delta_{\rho}\sqrt{Z\Delta_\rho}}\right)\omega^{(1)}+\right.\nonumber \\
& &\left.\frac{C_{\rho}Z_{\rho}^{\prime}}{2Z}\sqrt{\frac{\Delta_{\mu}}{Z}}\omega^{(3)}\right]+\left[\frac{C_{\mu}Z_{\mu}^{\prime}}{2Z}\sqrt{\frac{\Delta_{\rho}}{Z}}\omega^{(0)}+\frac{PC_{\mu}Z_{\mu}^{\prime}}{2Z\sqrt{Z\Delta_\rho}}\omega^{(1)}\right.-\nonumber \\
& &\left.\left(\frac{\Delta_{\mu}^{\prime}}{2\sqrt{Z\Delta_\mu}}+\frac{C_{\rho}Z_{\mu}^{\prime}}{2Z}\sqrt{\frac{\Delta_{\mu}}{Z}}\right)\omega^{(3)}\right]\wedge\nonumber \\
& &\left(-\frac{C_{\rho}Z_{\rho}^{\prime}}{2Z}\sqrt{\frac{\Delta_{\mu}}{z}}\omega^{(0)}-\frac{PC_{\mu}Z_{\mu}^{\prime}}{2Z\sqrt{Z\Delta_\rho}}\omega^{(2)}+\frac{C_{\mu}Z_{\rho}^{\prime}}{2Z}\sqrt{\frac{\Delta_{\rho}}{Z}}\omega^{(3)}\right) \\
\theta^{3}_{\;\;1}&=&-\theta^{1}_{\;\;3}=\left(\frac{-C_{\rho}Z_{\rho}^{\prime\prime}}{2Z^2}\sqrt{\Delta_{\rho}\Delta_{\mu}}+\frac{3C_{\rho}Z_{\rho}^{\prime}C_{\mu}Z_{\rho}^{\prime}}{4Z^3}\sqrt{\Delta_{\rho}\Delta_{\mu}}\right)\omega^{(1)}\wedge\omega^{(0)}-\nonumber \\
& &\left(\frac{C_{\rho}Z_{\rho}^{\prime}\Delta_{\mu}^{\prime}}{4Z^2}+\frac{3C_{\rho}Z_{\rho}^{\prime}C_{\rho}Z_{\rho}^{\prime}}{4Z^3}\Delta_{\mu}\right)\omega^{(2)}\wedge\omega^{(0)}-\nonumber \\
& &\frac{C_{\rho}Z_{\rho}^{\prime}}{2Z}\sqrt{\frac{\Delta_{\mu}}{Z}}\left[\left(\frac{\Delta_{\rho}^{\prime}}{2\sqrt{Z\Delta_\rho}}-\frac{C_{\mu}Z_{\rho}^{\prime}}{2Z}\sqrt{\frac{\Delta_{\rho}}{Z}}\right)\omega^{(1)}\wedge\omega^{(0)}\right.-\nonumber \\
& &\left.\frac{C_{\rho}Z_{\mu}^{\prime}}{2Z}\sqrt{\frac{\Delta_{\mu}}{Z}}\omega^{(2)}\wedge\omega^{(0)}+\frac{C_{\mu}Z_{\mu}^{\prime}}{Z}\sqrt{\frac{\Delta_{\rho}}{Z}}\omega^{(2)}\wedge\omega^{(3)}\right]+\nonumber \\
& &\left(\frac{C_{\mu}Z_{\rho}^{\prime\prime}\Delta_{\rho}}{2Z^2}+\frac{C_{\mu}Z_{\rho}^{\prime}\Delta_{\rho}^{\prime}}{4Z^2}-\frac{3(C_{\mu}Z_{\rho}^{\prime})^2\Delta_{\rho}}{4Z^3}\right)\omega^{(1)}\wedge\omega^{(3)}+ \nonumber \\
& &\frac{P}{\Delta_\rho}\left(\frac{C_{\mu}Z_{\rho}^{\prime\prime}\Delta_{\rho}}{2Z^2}+\frac{C_{\mu}Z_{\rho}^{\prime}\Delta_{\rho}^{\prime}}{4Z^2}-\frac{3(C_{\mu}Z_{\rho}^{\prime})^2\Delta_{\rho}}{4Z^3}\right)\omega^{(0)}\wedge\omega^{(3)}+ \nonumber \\
& &\frac{3C_{\mu}Z_{\rho}^{\prime}C_{\rho}Z_{\mu}^{\prime}}{4Z^3}\sqrt{\Delta_{\rho}\Delta_{\mu}}\omega^{(2)}\wedge\omega^{(3)}+\nonumber \\
& &\frac{C_{\mu}Z_{\rho}^{\prime}}{2Z}\sqrt{\frac{\Delta_{\rho}}{Z}}\left[\left(\frac{\Delta_{\mu}^{\prime}}{2\sqrt{Z\Delta_\mu}}+\frac{C_{\rho}Z_{\mu}^{\prime}}{2Z}\sqrt{\frac{\Delta_{\mu}}{Z}}\right)\omega^{(2)}\wedge\omega^{(3)}\right.+\nonumber \\
& &\left.\frac{C_{\mu}Z_{\rho}^{\prime}}{2Z}\sqrt{\frac{\Delta_{\rho}}{Z}}\omega^{(1)}\wedge\omega^{(3)}+\frac{PC_{\mu}Z_{\rho}^{\prime}}{2Z\sqrt{Z\Delta_{\rho}}}\omega^{(0)}\wedge\omega^{(3)}-\frac{C_{\rho}Z_{\rho}^{\prime}}{Z}\sqrt{\frac{\Delta_{\mu}}{Z}}\omega^{(1)}\wedge\omega^{(0)}\right]-\nonumber \\
& &\left(\frac{P^{\prime}C_{\mu}Z_{\mu}^{\prime}}{2Z^2}-\frac{PC_{\mu}Z_{\mu}^{\prime}\Delta_{\mu}^{\prime}}{4Z^{2}\Delta_{\rho}}-\frac{3C_{\mu}Z_{\mu}^{\prime}C_{\mu}Z_{\rho}^{\prime}P}{4Z^3}\right)\omega^{(1)}\wedge\omega^{(2)}- \nonumber \\
& &\frac{P}{\Delta_\rho}\left(\frac{P^{\prime}C_{\mu}Z_{\mu}^{\prime}}{2Z^2}-\frac{PC_{\mu}Z_{\mu}^{\prime}\Delta_{\rho}^{\prime}}{4Z^{2}\Delta_{\rho}}-\frac{3C_{\mu}Z_{\mu}^{\prime}C_{\mu}Z_{\rho}^{\prime}P}{4Z^3}\right)\omega^{(0)}\wedge\omega^{(3)}- \nonumber \\
& &\frac{PC_{\mu}Z_{\mu}^{\prime}}{2Z\sqrt{Z\Delta_{\rho}}}\left[\frac{C_{\mu}Z_{\rho}^{\prime}}{2Z}\sqrt{\frac{\Delta_{\rho}}{Z}}\omega^{(1)}\wedge\omega^{(2)}+\frac{PC_{\mu}Z_{\rho}^{\prime}}{2Z\sqrt{Z\Delta_{\rho}}}\omega^{(0)}\wedge\omega^{(2)}\right]+ \nonumber \\
& &\left[\frac{C_{\rho}Z_{\rho}^{\prime}}{2Z}\sqrt{\frac{\Delta_{\mu}}{Z}}\omega^{(1)}-\frac{C_{\mu}Z_{\mu}^{\prime}}{2Z}\sqrt{\frac{\Delta_{\rho}}{Z}}\omega^{(2)}+\frac{PC_{\mu}Z_{\rho}^{\prime}}{2Z\sqrt{Z\Delta_{\rho}}}\omega^{(3)}\right]\wedge \nonumber \\
& &\left[\left(\frac{\Delta_{\rho}^{\prime}}{2\sqrt{Z\Delta_\rho}}-\frac{C_{\mu}Z_{\rho}^{\prime}}{2Z}\sqrt{\frac{\Delta_{\rho}}{Z}}\right)\omega^{(0)}+\right. \nonumber \\
& &\left.\left(\frac{P^{\prime}}{\sqrt{Z\Delta_\rho}}-\frac{PC_{\mu}Z_{\rho}^{\prime}}{2Z\sqrt{Z\Delta_\rho}}-\frac{P\Delta_{\rho}^{\prime}}{2\Delta_{\rho}\sqrt{Z\Delta_\rho}}\right)\omega^{(1)}+\frac{C_{\rho}Z_{\rho}^{\prime}}{2Z}\sqrt{\frac{\Delta_{\mu}}{Z}}\omega^{(3)}\right]+ \nonumber \\
& &\left[\left(\frac{\Delta_{\mu}^{\prime}}{2\sqrt{Z\Delta_\mu}}+\frac{C_{\rho}Z_{\mu}^{\prime}}{2Z}\sqrt{\frac{\Delta_{\mu}}{Z}}\right)\omega^{(3)}-\frac{C_{\mu}Z_{\mu}^{\prime}}{2Z}\sqrt{\frac{\Delta_{\rho}}{Z}}\omega^{(0)}-\frac{PC_{\mu}Z_{\mu}^{\prime}}{2Z\sqrt{Z\Delta_{\rho}}}\omega^{(1)}\right]\wedge \nonumber \\
& &\left[\frac{C_{\rho}Z_{\mu}^{\prime}}{2Z}\sqrt{\frac{\Delta_{\mu}}{Z}}\omega^{(1)}+\frac{C_{\mu}Z_{\rho}^{\prime}}{2Z}\sqrt{\frac{\Delta_{\rho}}{Z}}\omega^{(2)}+\frac{PC_{\mu}Z_{\mu}^{\prime}}{2Z\sqrt{Z\Delta_{\rho}}}\omega^{(3)}\right] \\
\theta^{3}_{\;\;2}&=&-\theta^{2}_{\;\;3}=\left(\frac{\Delta_{\mu}^{\prime\prime}}{2Z}-\frac{\Delta_{\mu}^{\prime^{2}}}{4Z\Delta_{\mu}}+\frac{C_{\rho}Z_{\mu}^{\prime}\Delta_{\mu}^{\prime}}{2Z^2}+\frac{C_{\rho}Z_{\mu}^{\prime\prime}\Delta_{\mu}}{2Z^2}+\frac{3(C_{\rho}Z_{\mu}^{\prime})^2}{4Z^3}\Delta_{\mu}\right)\omega^{(2)}\wedge\omega^{(3)}- \nonumber \\
& &\left(\frac{\Delta_{\mu}^{\prime}C_{\mu}Z_{\rho}^{\prime}}{4Z^2}\sqrt{\frac{\Delta_{\rho}}{\Delta_{\mu}}}+\frac{3C_{\rho}Z_{\mu}^{\prime}C_{\mu}Z_{\rho}^{\prime}}{4Z^3}\sqrt{\Delta_{\rho}\Delta_{\mu}}\right)\omega^{(1)}\wedge\omega^{(3)}-\nonumber \\
& &P\left(\frac{\Delta_{\mu}^{\prime}C_{\mu}Z_{\rho}^{\prime}}{4Z^{2}\sqrt{\Delta_{\rho}\Delta_{\mu}}}+\frac{3C_{\rho}Z_{\mu}^{\prime}C_{\mu}Z_{\rho}^{\prime}}{4Z^3}\sqrt{\frac{\Delta_{\mu}}{\Delta_{\rho}}}\right)\omega^{(0)}\wedge\omega^{(3)}+\nonumber \\
& &\left(\frac{\Delta_{\mu}^{\prime}}{2\sqrt{Z\Delta_{\mu}}}+\frac{C_{\rho}Z_{\mu}^{\prime}}{2Z}\sqrt{\frac{\Delta_{\mu}}{Z}}\right)\left[\left(\frac{\Delta_{\mu}^{\prime}}{2\sqrt{Z\Delta_{\mu}}}+\frac{C_{\rho}Z_{\mu}^{\prime}}{2Z}\sqrt{\frac{\Delta_{\mu}}{Z}}\right)\omega^{(2)}\wedge\omega^{(3)}\right.+\nonumber \\
& &\left.\frac{C_{\mu}Z_{\rho}^{\prime}}{2Z}\sqrt{\frac{\Delta_{\rho}}{Z}}\omega^{(1)}\wedge\omega^{(3)}+\frac{PC_{\mu}Z_{\rho}^{\prime}}{2Z\sqrt{Z\Delta_{\rho}}}\omega^{(0)}\wedge\omega^{(3)}-\frac{C_{\rho}Z_{\rho}^{\prime}}{Z}\sqrt{\frac{\Delta_{\mu}}{Z}}\omega^{(1)}\wedge\omega^{(0)}\right]-\nonumber \\
& &\left(\frac{C_{\mu}Z_{\mu}^{\prime\prime}}{2Z^2}\sqrt{\Delta_{\rho}\Delta_{\mu}}+\frac{3C_{\mu}Z_{\mu}^{\prime}C_{\rho}Z_{\mu}^{\prime}}{4Z^3}\sqrt{\Delta_{\rho}\Delta_{\mu}}\right)\omega^{(2)}\wedge\omega^{(0)}+\nonumber \\
& &\left(-\frac{\Delta_{\rho}^{\prime}C_{\mu}Z_{\mu}^{\prime}}{4Z^2}+\frac{3C_{\mu}Z_{\mu}^{\prime}C_{\mu}Z_{\rho}^{\prime}}{4Z^3}\Delta_{\rho}\right)\omega^{(1)}\wedge\omega^{(0)}-\nonumber \\
& &\frac{C_{\mu}Z_{\mu}^{\prime}}{2Z}\sqrt{\frac{\Delta_{\rho}}{Z}}\left[\left(\frac{\Delta_{\rho}}{2\sqrt{Z\Delta_{\rho}}}-\frac{C_{\mu}Z_{\rho}^{\prime}}{2Z}\sqrt{\frac{\Delta_{\rho}}{Z}}\right)\omega^{(1)}\wedge\omega^{(0)}-\frac{C_{\rho}Z_{\mu}^{\prime}}{2Z}\sqrt{\frac{\Delta_{\mu}}{Z}}\omega^{(2)}\wedge\omega^{(0)}+\right.\nonumber \\
& &\left.\frac{C_{\mu}Z_{\mu}^{\prime}}{Z}\sqrt{\frac{\Delta_{\rho}}{Z}}\omega^{(2)}\wedge\omega^{(3)}\right]+P\left(-\frac{P^{\prime}C_{\mu}Z_{\mu}^{\prime}}{2Z^{2}\Delta_{\rho}}+\frac{PC_{\mu}Z_{\mu}^{\prime}\Delta_{\rho}^{\prime}}{4Z^{2}\Delta_{\rho}^{2}}+\right.\nonumber \\
& &\left.\frac{3PC_{\mu}Z_{\mu}^{\prime}C_{\mu}Z_{\rho}^{\prime}}{4Z^{3}\Delta_{\rho}}\right)\omega^{(0)}\wedge\omega^{(1)}-\nonumber \\
& &\left(\frac{PC_{\mu}Z_{\mu}^{\prime\prime}}{2Z^2}\sqrt{\frac{\Delta_\mu}{\Delta_\rho}}+\frac{3PC_{\mu}Z_{\mu}^{\prime}C_{\rho}Z_{\mu}^{\prime}}{4Z^{3}}\sqrt{\frac{\Delta_\mu}{\Delta_\rho}}\right)\omega^{(2)}\wedge\omega^{(1)}+\nonumber \\
& &\frac{PC_{\mu}Z_{\mu}^{\prime}}{2Z\sqrt{Z\Delta_{\rho}}}\left[\frac{C_{\rho}Z_{\mu}^{\prime}}{2Z}\sqrt{\frac{\Delta_{\mu}}{Z}}\omega^{(2)}\wedge\omega^{(1)}+\right.\nonumber \\
& &\left.\left(\frac{P^{\prime}}{\sqrt{Z\Delta_\rho}}-\frac{PC_{\mu}Z_{\rho}^{\prime}}{2Z\sqrt{Z\Delta_{\rho}}}-\frac{P\Delta_{\rho}^{\prime}}{2\Delta_{\rho}\sqrt{Z\Delta_\rho}}\right)\right.\omega^{(1)}\wedge\omega^{(0)}\nonumber \\
& &\left.\frac{PC_{\mu}Z_{\mu}^{\prime}}{Z\sqrt{Z\Delta_{\rho}}}\omega^{(2)}\wedge\omega^{(3)}\right]+\nonumber \\
& &\left(\frac{C_{\rho}Z_{\rho}^{\prime}}{2Z}\sqrt{\frac{\Delta_{\mu}}{Z}}\omega^{(1)}-\frac{C_{\mu}Z_{\mu}^{\prime}}{2Z}\sqrt{\frac{\Delta_{\rho}}{Z}}\omega^{(2)}+\frac{PC_{\mu}Z_{\rho}^{\prime}}{2Z\sqrt{Z\Delta_{\rho}}}\omega^{(3)}\right)\wedge \nonumber \\
& &\left(-\frac{C_{\rho}Z_{\mu}^{\prime}}{2Z}\sqrt{\frac{\Delta_{\mu}}{Z}}\omega^{(0)}+\frac{PC_{\mu}Z_{\rho}^{\prime}}{2Z\sqrt{Z\Delta_{\rho}}}\omega^{(2)}+\frac{C_{\mu}Z_{\mu}^{\prime}}{2Z}\sqrt{\frac{\Delta_{\rho}}{Z}}\omega^{(3)}\right)+\nonumber \\
& &\left[-\frac{C_{\rho}Z_{\rho}^{\prime}}{2Z}\sqrt{\frac{\Delta_{\mu}}{Z}}\omega^{(0)}-\frac{PC_{\mu}Z_{\mu}^{\prime}}{2Z\sqrt{Z\Delta_{\rho}}}\omega^{(2)}+\frac{C_{\mu}Z_{\rho}^{\prime}}{2Z}\sqrt{\frac{\Delta_{\rho}}{Z}}\omega^{(3)}\right]\wedge\nonumber \\
& &\left[-\frac{C_{\rho}Z_{\mu}^{\prime}}{2Z}\sqrt{\frac{\Delta_{\mu}}{Z}}\omega^{(1)}-\frac{C_{\mu}Z_{\rho}^{\prime}}{2Z}\sqrt{\frac{\Delta_{\rho}}{Z}}\omega^{(2)}-\frac{PC_{\mu}Z_{\mu}^{\prime}}{2Z\sqrt{Z\Delta_{\rho}}}\omega^{(3)}\right]
\end{eqnarray}
\section{Riemann and Ricci Curvature Tensors}
The "00" componentof the Ricci tensor is equal to
\begin{equation}
R_{00}=R^{1}_{010}+R^{2}_{020}+R^{3}_{030}
\end{equation}
According to (45) and (46) and by considering the coefficient of $\omega^{(1)}\wedge\omega^{(0)}$ in the $\theta_{\;\;0}^{1}$, $R^{1}_{010}$ will be
\begin{eqnarray}
R^{1}_{010}&=&\frac{\Delta_{\rho}^{\prime\prime}}{2Z}-\frac{C_{\mu}Z_{\rho}^{\prime}\Delta_{\rho}^{\prime}}{Z^2}-\frac{C_{\mu}Z_{\rho}^{\prime\prime}\Delta_{\rho}}{2Z^2}+\frac{(C_{\mu}Z_{\rho}^{\prime})^{2}\Delta_{\rho}}{Z^3}-\frac{3(C_{\rho}Z_{\rho}^{\prime})^{2}\Delta_{\mu}}{4Z^3}+\nonumber \\
& &\frac{(C_{\rho}Z_{\mu}^{\prime})^{2}\Delta_{\mu}}{4Z^3}-\frac{PP^{\prime\prime}}{Z\Delta_{\rho}}+\frac{2PP^{\prime}\Delta_{\rho}^{\prime}}{Z\Delta_{\rho}^{2}}+\frac{P^{2}C_{\mu}Z_{\rho}^{\prime\prime}}{2Z^{2}\Delta_{\rho}}-\frac{P^{2}(C_{\mu}Z_{\rho}^{\prime})^2}{Z^{3}\Delta_{\rho}}-\nonumber \\
& &\frac{P^{2}C_{\mu}Z_{\rho}^{\prime}\Delta_{\rho}^{\prime}}{Z^{2}\Delta_{\rho}^{2}}+\frac{P^{2}\Delta_{\rho}^{\prime\prime}}{2Z\Delta_{\rho}^{2}}-\frac{P^{2}\Delta_{\rho}^{\prime^{2}}}{Z\Delta_{\rho}^{3}}-\frac{P^{\prime^{2}}}{Z\Delta_{\rho}}+\frac{2PP^{\prime}C_{\mu}Z_{\rho}^{\prime}}{Z^{2}\Delta_{\rho}}
\end{eqnarray}
$R^{2}_{020}$ comes from the coefficient of $\omega^{(2)}\wedge\omega^{(0)}$ in the $\theta_{\;\;0}^{2}$ in (47),
\begin{eqnarray}
R^{2}_{020}&=&-\frac{C_{\rho}Z_{\mu}^{\prime}\Delta_{\mu}^{\prime}}{4Z^2}-\frac{C_{\rho}Z_{\mu}^{\prime}\Delta_{\mu}}{2Z^2}-\frac{(C_{\rho}Z_{\mu}^{\prime})^{2}\Delta_{\mu}}{2Z^3}+\frac{(C_{\mu}Z_{\rho}^{\prime})^{2}P^2}{2Z^3\Delta_{\rho}}-\nonumber \\
& &\frac{C_{\mu}Z_{\rho}^{\prime\prime}P^2}{4Z^{2}\Delta_{\rho}}-\frac{C_{\mu}Z_{\rho}^{\prime}PP^{\prime}}{2Z^{2}\Delta_{\rho}}+\frac{C_{\mu}Z_{\rho}^{\prime}P^{2}\Delta_{\rho}^{\prime}}{4Z^{2}\Delta_{\rho}^{2}}+\frac{C_{\mu}Z_{\rho}^{\prime}\Delta_{\rho}^{\prime}}{4Z^2}-\nonumber \\
& &\frac{(C_{\mu}Z_{\rho}^{\prime})^{2}\Delta_{\rho}}{4Z^3}+\frac{(C_{\mu}Z_{\mu}^{\prime})^{2}\Delta_{\rho}}{4Z^3} 
\end{eqnarray}
and the coefficient of $\omega^{(3)}\wedge\omega^{(0)}$ in the $\theta_{\;\;0}^{3}$ ,(48) ,gives  $R^{3}_{030}$ as
\begin{eqnarray}
R^{3}_{030}&=&-\frac{C_{\mu}Z_{\rho}^{\prime}PP^{\prime}}{2Z^{2}\Delta_{\rho}}+\frac{C_{\mu}Z_{\rho}^{\prime}P^{2}\Delta_{\rho}^{\prime}}{4Z^{2}\Delta_{\rho}^{2}}+\frac{(C_{\mu}Z_{\rho}^{\prime})^{2}P^2}{2Z^3\Delta_{\rho}}-\frac{C_{\mu}Z_{\rho}^{\prime\prime}P^2}{2Z^2\Delta_{\rho}}+\nonumber \\
& &\frac{C_{\mu}Z_{\rho}^{\prime}\Delta_{\rho}^{\prime}}{4Z^2}-\frac{(C_{\mu}Z_{\rho}^{\prime})^{2}\Delta_{\rho}}{4Z^3}+\frac{(C_{\rho}Z_{\rho}^{\prime})^{2}\Delta_{\mu}}{4Z^3}-\frac{C_{\rho}Z_{\mu}^{\prime}\Delta_{\mu}^{\prime}}{4Z^2}-\nonumber \\
& &\frac{(C_{\rho}Z_{\mu}^{\prime})^{2}\Delta_{\mu}}{4Z^3}+\frac{(C_{\mu}Z_{\mu}^{\prime})^{2}\Delta_{\rho}}{4Z^3}
\end{eqnarray}
Inserting (53)-(55)in (52) gives
\begin{eqnarray}
R_{00}&=&\frac{\Delta_{\rho}^{\prime\prime}}{2Z}-\frac{C_{\mu}Z_{\rho}^{\prime}\Delta_{\rho}^{\prime}}{2Z^2}-\frac{C_{\mu}Z_{\rho}^{\prime\prime}\Delta_{\rho}}{2Z^2}+\frac{(C_{\mu}Z_{\rho}^{\prime})^{2}\Delta_{\rho}}{2Z^3}-\frac{(C_{\rho}Z_{\rho}^{\prime})^{2}\Delta_{\mu}}{2Z^3}-\frac{(C_{\rho}Z_{\mu}^{\prime})^{2}\Delta_{\mu}}{2Z^3}\nonumber \\
& &-\frac{C_{\rho}Z_{\mu}^{\prime}\Delta_{\mu}}{2Z^2}-\frac{C_{\rho}Z_{\mu}^{\prime}\Delta_{\mu}^{\prime}}{2Z^2}+\frac{(C_{\mu}Z_{\mu}^{\prime})^{2}\Delta_{\rho}}{2Z^3}-\frac{PP^{\prime\prime}}{Z\Delta_{\rho}}+\frac{2PP^{\prime}\Delta_{\rho}^{\prime}}{Z\Delta_{\rho}^{2}}-\nonumber \\
& &\frac{C_{\mu}Z_{\rho}^{\prime\prime}P^2}{2Z^2\Delta_{\rho}}-\frac{C_{\mu}Z_{\rho}^{\prime}P^{2}\Delta_{\rho}^{\prime}}{4Z^{2}\Delta_{\rho}^{2}}+\frac{P^{2}\Delta_{\rho}^{\prime\prime}}{2Z\Delta_{\rho}^{2}}-\frac{P^{2}\Delta_{\rho}^{\prime^{2}}}{Z\Delta_{\rho}^{3}}-\frac{P^{\prime^{2}}}{Z\Delta_{\rho}}+\frac{PP^{\prime}C_{\mu}Z_{\rho}^{\prime}}{Z^{2}\Delta_{\rho}}
\end{eqnarray}
Next, the "01" component of Ricci tensor is
\begin{equation}
R_{01}=R^{2}_{021}+R^{3}_{031}
\end{equation}
where $R^{2}_{021}$ comes from the $\omega^{(2)}\wedge\omega^{(1)}$ factor of $\theta_{\;\;0}^{2}$:
\begin{equation}
R^{2}_{021}=-\frac{C_{\mu}Z_{\rho}^{\prime\prime}P}{2Z^2}+\frac{(C_{\mu}Z_{\mu}^{\prime})^{2}P}{4Z^3}+\frac{(C_{\mu}Z_{\rho}^{\prime})^{2}P}{4Z^3}
\end{equation}
and $R^{3}_{031}$ is equal to the $\omega^{(3)}\wedge\omega^{(1)}$ factor of $\theta_{\;\;0}^{3}$:
\begin{equation}
R^{3}_{031}=-\frac{C_{\mu}Z_{\rho}^{\prime\prime}P}{2Z^2}+\frac{(C_{\mu}Z_{\rho}^{\prime})^{2}P}{4Z^3}+\frac{(C_{\mu}Z_{\mu}^{\prime})^{2}P}{4Z^3}
\end{equation}
Thus from (57)-(59) $R_{01}$ is
\begin{equation}
R_{01}=-\frac{C_{\mu}Z_{\rho}^{\prime\prime}P}{Z^2}+\frac{(C_{\mu}Z_{\mu}^{\prime})^{2}P}{2Z^3}+\frac{(C_{\mu}Z_{\rho}^{\prime})^{2}P}{2Z^3}
\end{equation}
$R_{02}$ is equal to
\begin{equation}
R_{02}=R^{1}_{012}+R^{3}_{032}
\end{equation}
where $R^{1}_{012}$ comes from the $\omega^{(1)}\wedge\omega^{(2)}$ factor of $\theta_{\;\;0}^{1}$ in (46) and $R^{3}_{032}$ the $\omega^{(3)}\wedge\omega^{(2)}$ factor of $\theta_{\;\;0}^{3}$ in (48). Since both are zero, we have
\begin{equation}
R_{02}=0
\end{equation}
The "03" component of Ricci tensor $R_{03}$ is 
\begin{equation}
R_{03}=R^{1}_{013}+R^{2}_{023}
\end{equation}
while $R^{1}_{013}$ is the $\omega^{(1)}\wedge\omega^{(3)}$ factor of $\theta_{\;\;0}^{1}$ in (46)
\begin{equation}
R^{1}_{013}=\frac{C_{\rho}Z_{\rho}^{\prime\prime}}{2Z^2}\sqrt{\Delta_{\rho}\Delta_{\mu}}-\frac{C_{\rho}Z_{\rho}^{\prime}C_{\mu}Z_{\rho}^{\prime}}{4Z^3}\sqrt{\Delta_{\rho}\Delta_{\mu}}-\frac{C_{\rho}Z_{\mu}^{\prime}C_{\mu}Z_{\mu}^{\prime}}{4Z^3}\sqrt{\Delta_{\rho}\Delta_{\mu}}
\end{equation}
and $R^{2}_{023}$ is the $\omega^{(2)}\wedge\omega^{(3)}$ factor of $\theta_{\;\;0}^{2}$ in (47)
\begin{equation}
R^{2}_{023}=\frac{C_{\mu}Z_{\mu}^{\prime\prime}}{2Z^2}\sqrt{\Delta_{\rho}\Delta_{\mu}}+\frac{C_{\rho}Z_{\rho}^{\prime}C_{\mu}Z_{\rho}^{\prime}}{4Z^3}\sqrt{\Delta_{\rho}\Delta_{\mu}}+\frac{C_{\rho}Z_{\mu}^{\prime}C_{\mu}Z_{\mu}^{\prime}}{4Z^3}\sqrt{\Delta_{\rho}\Delta_{\mu}}
\end{equation}
Putting (64) and (65) in (63)
\begin{equation}
R_{03}=\frac{C_{\rho}Z_{\rho}^{\prime\prime}}{2Z^2}\sqrt{\Delta_{\rho}\Delta_{\mu}}+\frac{C_{\mu}Z_{\mu}^{\prime\prime}}{2Z^2}\sqrt{\Delta_{\rho}\Delta_{\mu}}
\end{equation}
Next diagonal component of Ricci tensor ,"11" ,is
\begin{equation}
R_{11}=R^{0}_{101}+R^{2}_{121}+R^{3}_{131}
\end{equation}
where
\begin{equation}
R^{0}_{101}=-R^{1}_{010}
\end{equation}
and $R^{2}_{121}$ comes from the $\omega^{(2)}\wedge\omega^{(1)}$ factor of $\theta_{\;\;1}^{2}$ in (49)
\begin{eqnarray}
R^{2}_{121}&=&\frac{C_{\rho}Z_{\mu}^{\prime\prime}\Delta_{\mu}}{2Z^2}+\frac{(C_{\rho}Z_{\mu}^{\prime})^{2}\Delta_{\mu}}{2Z^3}-\frac{C_{\mu}Z_{\rho}^{\prime\prime}\Delta_{\rho}}{2Z^2}-\frac{C_{\mu}Z_{\rho}^{\prime}\Delta_{\rho}^{\prime}}{4Z^2}+\frac{(C_{\mu}Z_{\rho}^{\prime})^{2}\Delta_{\rho}}{2Z^3}+\nonumber \\
& &\frac{C_{\rho}Z_{\mu}^{\prime}\Delta_{\mu}^{\prime}}{4Z^2}+\frac{C_{\mu}Z_{\rho}^{\prime}PP^{\prime}}{2Z^{2}\Delta_{\rho}}-\frac{(C_{\mu}Z_{\rho}^{\prime})^{2}P^2}{4Z^3\Delta_{\rho}}-\frac{C_{\mu}Z_{\rho}^{\prime}\Delta_{\rho}^{\prime}P^2}{4Z^{2}\Delta_{\rho}^2}+\nonumber \\
& &\frac{(C_{\mu}Z_{\mu}^{\prime})^{2}P^2}{4Z^3\Delta_{\rho}}
\end{eqnarray}
$R^{3}_{131}$ may be obtained by considering the coefficient of $\omega^{(3)}\wedge\omega^{(1)}$ in $\theta_{\;\;1}^{3}$ in (50)
\begin{eqnarray}
R^{3}_{131}&=&-\frac{C_{\mu}Z_{\rho}^{\prime\prime}\Delta_{\rho}}{2Z^2}-\frac{C_{\mu}Z_{\rho}^{\prime}\Delta_{\rho}^{\prime}}{4Z^2}+\frac{(C_{\mu}Z_{\rho}^{\prime})^{2}\Delta_{\rho}}{2Z^3}+\frac{C_{\rho}Z_{\mu}^{\prime}\Delta_{\mu}^{\prime}}{4Z^2}-\nonumber \\
& &\frac{(C_{\rho}Z_{\rho}^{\prime})^{2}\Delta_{\mu}}{4Z^3}+\frac{(C_{\rho}Z_{\mu}^{\prime})^{2}\Delta_{\mu}}{4Z^3}+\frac{C_{\mu}Z_{\rho}^{\prime}PP^{\prime}}{2Z^{2}\Delta_{\rho}}-\frac{(C_{\mu}Z_{\rho}^{\prime})^{2}P^2}{4Z^3\Delta_{\rho}}-\nonumber \\
& &\frac{C_{\mu}Z_{\rho}^{\prime}\Delta_{\rho}^{\prime}P^2}{4Z^{2}\Delta_{\rho}^2}+\frac{(C_{\mu}Z_{\mu}^{\prime})^{2}P^2}{4Z^3\Delta_{\rho}}
\end{eqnarray}
Inserting (53) and (68)-(70) in (67) gives
\begin{eqnarray}
R_{11}&=&-\frac{\Delta_{\rho}^{\prime\prime}}{2Z}+\frac{C_{\mu}Z_{\rho}^{\prime}\Delta_{\rho}^{\prime}}{2Z^2}+\frac{(C_{\rho}Z_{\rho}^{\prime})^{2}\Delta_{\mu}}{2Z^3}+\frac{(C_{\rho}Z_{\mu}^{\prime})^{2}\Delta_{\mu}}{2Z^3}+\frac{C_{\rho}Z_{\mu}^{\prime\prime}\Delta_{\mu}}{2Z^2}+\frac{C_{\rho}Z_{\mu}^{\prime}\Delta_{\mu}^{\prime}}{2Z^2}\nonumber \\
& &-\frac{C_{\mu}Z_{\rho}^{\prime\prime}\Delta_{\rho}}{2Z^2}+\frac{PP^{\prime\prime}}{Z\Delta_{\rho}}-\frac{2PP^{\prime}\Delta^{\prime}}{Z\Delta_{\rho}^{2}}-\frac{C_{\mu}Z_{\rho}^{\prime\prime}P^2}{2Z^2\Delta_{\rho}}+\frac{(C_{\mu}Z_{\mu}^{\prime})^{2}P^2}{2Z^{3}\Delta_{\rho}}+\frac{C_{\mu}Z_{\rho}^{\prime}P^{2}\Delta_{\rho}^{\prime}}{2Z^{2}\Delta_{\rho}^{2}}\nonumber \\
& &-\frac{P^{2}\Delta_{\rho}^{\prime\prime}}{2Z\Delta_{\rho}^{2}}+\frac{P^{2}\Delta_{\rho}^{\prime^{2}}}{Z\Delta_{\rho}^{3}}+\frac{P^{\prime^{2}}}{Z\Delta_{\rho}}-\frac{C_{\mu}Z_{\rho}^{\prime}PP^\prime}{Z^{2}\Delta_{\rho}}+\frac{(C_{\mu}Z_{\rho}^{\prime})^{2}P^2}{2Z^{3}\Delta_{\rho}}
\end{eqnarray}
Also we have
$R_{12}=R^{0}_{102}+R^{3}_{132}$,
where $R^{0}_{102}$ comes from the $\omega^{(0)}\wedge\omega^{(2)}$ factor of $\theta_{1}^{0}$ and $R^{3}_{132}$ from the $\omega^{(3)}\wedge\omega^{(2)}$ factor of $\theta_{\;\;1}^{3}$ .Both of these factors are zero so we have 
\begin{equation}
R_{12}=0
\end{equation}
The "13" component of Ricci tensor is
\begin{equation}
R_{13}=R^{0}_{103}+R^{2}_{123}
\end{equation}
where $R^{0}_{103}$ is the $\omega^{(0)}\wedge\omega^{(3)}$ factor of $\theta_{\;\;1}^{0}$ and $R^{2}_{123}$ comes from the $\omega^{(2)}\wedge\omega^{(3)}$ factor of $\theta_{\;\;1}^{2}$ 
\begin{eqnarray}
R^{0}_{103}&=&P\left(\frac{C_{\rho}Z_{\rho}^{\prime\prime}}{2Z^2}\sqrt{\frac{\Delta_\mu}{\Delta_\rho}}-\frac{C_{\rho}Z_{\rho}^{\prime}C_{\mu}Z_{\rho}^{\prime}}{4Z^3}\sqrt{\frac{\Delta_\mu}{\Delta_\rho}}-\frac{C_{\mu}Z_{\mu}^{\prime}C_{\rho}Z_{\mu}^{\prime}}{4Z^3}\sqrt{\frac{\Delta_\mu}{\Delta_\rho}}\right) \\
R^{2}_{123}&=&P\left(\frac{C_{\mu}Z_{\mu}^{\prime\prime}}{2Z^2}\sqrt{\frac{\Delta_\mu}{\Delta_\rho}}+\frac{C_{\mu}Z_{\mu}^{\prime}C_{\rho}Z_{\mu}^{\prime}}{4Z^3}\sqrt{\frac{\Delta_\mu}{\Delta_\rho}}+\frac{C_{\mu}Z_{\rho}^{\prime}C_{\rho}Z_{\rho}^{\prime}}{4Z^3}\sqrt{\frac{\Delta_\mu}{\Delta_\rho}}\right) 
\end{eqnarray}
Now from (73)-(75) we simply obtain
\begin{equation}
R_{13}=\frac{P}{2Z^2}(C_{\rho}Z_{\rho}^{\prime\prime}+C_{\mu}Z_{\mu}^{\prime\prime})\sqrt{\frac{\Delta_\mu}{\Delta_\rho}} 
\end{equation}
The other diagonal component is $R_{22}$:
\begin{equation}
R_{22}=R^{0}_{202}+R^{1}_{212}+R^{3}_{232}
\end{equation}
Also we know
\begin{equation} 
R^{0}_{202}=-R^{2}_{020}\hspace{0.7cm} , \hspace{0.7cm} R^{1}_{212}=R^{2}_{121}
\end{equation}
The remaining term in (77) is $R^{3}_{232}$ which can be obtained from the coefficient of  $\omega^{(3)}\wedge\omega^{(2)}$ in $\theta_{\;\;2}^{3}$ from (51)
\begin{eqnarray}
R^{3}_{232}&=&-\frac{\Delta_{\mu}^{\prime\prime}}{2Z}-\frac{C_{\rho}Z_{\mu}^{\prime\prime}\Delta_{\mu}}{2Z^2}-\frac{(C_{\rho}Z_{\mu}^{\prime})^{2}\Delta_{\mu}}{Z^3}-\frac{3(C_{\mu}Z_{\mu}^{\prime})^{2}\Delta_{\rho}}{4Z^3}-\frac{(C_{\mu}Z_{\rho}^{\prime})^{2}\Delta_{\rho}}{4Z^3}\nonumber \\
& &-\frac{C_{\rho}Z_{\mu}^{\prime}\Delta_{\mu}^{\prime}}{Z^2}-\frac{3(C_{\mu}Z_{\rho}^{\prime})^{2}P^{2}}{4Z^{3}\Delta_{\rho}}+\frac{(C_{\mu}Z_{\rho}^{\prime})^{2}P^{2}}{4Z^{3}\Delta_{\rho}}
\end{eqnarray}
Putting (54),(69),(78),and (79) in (77) leads to
\begin{eqnarray}
R_{22}&=&\frac{C_{\rho}Z_{\mu}^{\prime\prime}\Delta_{\mu}}{2Z^2}-\frac{C_{\rho}Z_{\mu}^{\prime}\Delta_{\mu}^{\prime}}{Z^2}-\frac{C_{\mu}Z_{\rho}^{\prime\prime}\Delta_{\rho}}{2Z^2}-\frac{C_{\mu}Z_{\rho}^{\prime}\Delta_{\rho}^{\prime}}{2Z^2}+\frac{(C_{\mu}Z_{\rho}^{\prime})^{2}\Delta_{\rho}}{4Z^3}\nonumber \\
& &-\frac{\Delta_{\mu}^{\prime\prime}}{2Z}+\frac{(C_{\mu}Z_{\mu}^{\prime})^{2}\Delta_{\rho}}{2Z^3}-\frac{(C_{\mu}Z_{\rho}^{\prime})^{2}P^{2}}{2Z^{3}\Delta_{\rho}}+\frac{C_{\mu}Z_{\rho}^{\prime\prime}P^2}{2Z^{2}\Delta_{\rho}}+\frac{C_{\mu}Z_{\rho}^{\prime}PP^{\prime}}{Z^{2}\Delta_{\rho}}\nonumber \\
& &-\frac{C_{\mu}Z_{\rho}^{\prime}\Delta_{\rho}^{\prime}P^2}{2Z^{2}\Delta_{\rho}^{2}}-\frac{(C_{\mu}Z_{\mu}^{\prime})^{2}P^{2}}{2Z^{3}\Delta_{\rho}}
\end{eqnarray}
Last components of the Ricci tensor remained are $R_{23}$ and $R_{33}$. The off-diagonal one, $R_{23}$ is defined as 
\begin{equation}
R_{23}=R^{0}_{203}+R^{1}_{213}
\end{equation}
in which $R^{0}_{203}$ comes from the $\omega^{(0)}\wedge\omega^{(3)}$ factor of $\theta_{\;\;2}^{0}$ and $R^{1}_{213}$ from the $\omega^{(1)}\wedge\omega^{(3)}$ factor of $\theta_{\;\;2}^{1}$ .Both of these factors are zero and we have 
\begin{equation}
R_{23}=0
\end{equation}
Finally the remaining diagonal component is
\begin{equation}
R_{33}=R^{0}_{303}+R^{1}_{313}+R^{2}_{323}
\end{equation}
where we have
\begin{equation}
R^{0}_{303}=-R^{0}_{030}\;\; , \;\;R^{1}_{313}=R^{3}_{131}\;\; , \;\;R^{2}_{323}=R^{3}_{232}
\end{equation}
Inserting (55),(70),(79),and (84) in (83) we get
\begin{eqnarray}
R_{33}&=&-\frac{C_{\mu}Z_{\rho}^{\prime}\Delta_{\rho}^{\prime}}{2Z^2}-\frac{C_{\rho}Z_{\mu}^{\prime}\Delta_{\mu}^{\prime}}{2Z^2}-\frac{C_{\mu}Z_{\rho}^{\prime\prime}\Delta_{\rho}}{2Z^2}-\frac{C_{\rho}Z_{\mu}^{\prime\prime}\Delta_{\mu}}{2Z^2}-\frac{\Delta_{\mu}^{\prime\prime}}{2Z}\nonumber \\
& &+\frac{(C_{\mu}Z_{\rho}^{\prime})^{2}\Delta_{\rho}}{2Z^3}-\frac{(C_{\rho}Z_{\rho}^{\prime})^{2}\Delta_{\mu}}{2Z^3}-\frac{(C_{\rho}Z_{\mu}^{\prime})^{2}\Delta_{\mu}}{2Z^3}+\frac{(C_{\mu}Z_{\mu}^{\prime})^{2}\Delta_{\rho}}{2Z^3}+\frac{C_{\mu}Z_{\rho}^{\prime}PP^{\prime}}{Z^{2}\Delta_{\rho}}\nonumber \\
& &-\frac{C_{\mu}Z_{\rho}^{\prime}\Delta_{\rho}^{\prime}P^2}{2Z^{2}\Delta_{\rho}^{2}}-\frac{(C_{\mu}Z_{\rho}^{\prime})^{2}P^{2}}{2Z^{3}\Delta_{\rho}}-\frac{(C_{\mu}Z_{\mu}^{\prime})^{2}P^{2}}{2Z^{3}\Delta_{\rho}}+\frac{C_{\mu}Z_{\rho}^{\prime\prime}P^2}{2Z^{2}\Delta_{\rho}}
\end{eqnarray}
It is easily obtained from (56),(71),(80),and (85) that
\begin{equation}
R_{00}+R_{11}=\left(\frac{(C_{\mu}Z_{\rho}^{\prime})^{2}}{2Z^3}+\frac{(C_{\mu}Z_{\mu}^{\prime})^{2}}{2Z^3}-\frac{C_{\mu}Z_{\rho}^{\prime\prime}}{Z^2}\right)\left(\Delta_{\rho}+\frac{P^2}{\Delta_{\rho}}\right)
\end{equation}
and
\begin{equation}
R_{22}-R_{33}=\frac{(C_{\rho}Z_{\rho}^{\prime})^{2}\Delta_{\mu}}{2Z^3}+\frac{(C_{\rho}Z_{\mu}^{\prime})^{2}}{2Z^3}-\frac{C_{\rho}Z_{\mu}^{\prime\prime}}{Z^2}\Delta_{\mu}
\end{equation}
Einstein vacuum field equation can be written as
\begin{equation}
R_{\mu\nu}=\Lambda\eta_{\mu\nu}
\end{equation}
From the "03" component of the field equations (88), and (66) we can infere that $Z_{\mu}$ and $Z_{\rho}$ are quadratic functions of $\mu$ and $\rho$ ,respectively.
In applying the suitable boundary conditions and in comparison with Carter's results[12] we may choose
\begin{eqnarray}
Z_{\rho}&=&A(\rho^{2}+a^2) \\
Z_{\mu}&=&Aa(1-\mu^2) \\
C_{\mu}&=&A \\
C_{\rho}&=&aA
\end{eqnarray}
where $A$ and $a$ are constants.(89) and (90) imply all of the off-diagonal components of the field equations to be satisfied automatically. Also this leads to the vanishing of the right hand side of (86) and (87) which means that all the diagonal components of field equations are not independent of each other.
From the fact that $Z_{\rho}$ and $Z_{\mu}$ are quadratic and using (60),(86),(87) in (71) and (78), $R_{11}$ and $R_{22}$ take the simpler form
\begin{eqnarray}
R_{11}&=&-\frac{\Delta_{\rho}^{\prime\prime}}{2Z}+\frac{C_{\mu}Z_{\rho}^{\prime}\Delta_{\rho}^{\prime}}{2Z^2}-\frac{C_{\rho}Z_{\mu}^{\prime\prime}\Delta_{\mu}}{2Z^2}+\frac{C_{\rho}Z_{\mu}^{\prime}\Delta_{\mu}^{\prime}}{Z^2}-\frac{C_{\mu}Z_{\rho}^{\prime\prime}\Delta_{\mu}}{2Z^2}+\frac{PP^{\prime\prime}}{Z\Delta_{\rho}}\nonumber \\
& &-\frac{2PP^{\prime}\Delta_{\rho}^{\prime}}{2\Delta_{\rho}^{2}}+\frac{C_{\mu}Z_{\rho}^{\prime\prime}P^2}{2Z^{2}\Delta_{\rho}}-\frac{P^{2}\Delta_{\rho}^{\prime\prime}}{2Z\Delta_{\rho}}+\frac{P^{2}\Delta_{\rho}^{\prime^{2}}}{Z\Delta_{\rho}^{3}}+\frac{P^{\prime^{2}}}{Z\Delta_{\rho}}\nonumber \\
& &-\frac{C_{\mu}Z_{\rho}^{\prime}PP^{\prime}}{Z^{2}\Delta_{\rho}}+\frac{C_{\mu}Z_{\rho}^{\prime}\Delta_{\rho}^{\prime}P^2}{2Z^{2}\Delta_{\rho}^{2}} \\
R_{22}&=&\frac{C_{\rho}Z_{\mu}^{\prime\prime}\Delta_{\mu}}{2Z^2}-\frac{C_{\rho}Z_{\mu}^{\prime}\Delta_{\mu}^{\prime}}{2Z^2}-\frac{C_{\mu}Z_{\rho}^{\prime}\Delta_{\mu}^{\prime}}{2Z^2}+\frac{C_{\rho}Z_{\mu}^{\prime\prime}\Delta_{\rho}}{2Z^2}-\frac{\Delta_{\mu}^{\prime\prime}}{2Z}\nonumber \\
& &-\frac{C_{\mu}Z_{\rho}^{\prime\prime}P^2}{2Z^{2}\Delta_{\rho}}+\frac{C_{\mu}Z_{\rho}^{\prime}PP^{\prime}}{Z^{2}\Delta_{\rho}}-\frac{C_{\mu}Z_{\rho}^{\prime}\Delta_{\rho}^{\prime}P^2}{2Z^{2}\Delta_{\rho}^{2}}
\end{eqnarray}
From (88) we have 
\begin{equation}
R_{11}+R_{22}=2\Lambda
\end{equation}
Putting (93) and (94) in (95) gives
\begin{equation}
-\frac{\Delta_{\rho}^{\prime\prime}}{2Z}-\frac{\Delta_{\mu}^{\prime\prime}}{2Z}+\frac{PP^{\prime\prime}}{Z\Delta_{\rho}}-\frac{2PP^{\prime}\Delta_{\rho}^{\prime}}{Z\Delta_{\rho}^{2}}-\frac{P^{2}\Delta_{\rho}^{\prime\prime}}{2Z\Delta_{\rho}}+\frac{P^{2}\Delta_{\rho}^{\prime^{2}}}{Z\Delta_{\rho}^{3}}+\frac{P^{\prime^{2}}}{Z\Delta_{\rho}}=2\Lambda
\end{equation}
From (89)-(92) and (19) we have
\begin{equation}
Z=A^{2}(\rho^{2}+a^{2}\mu^{2})
\end{equation}
Multiplying both sides of (96) by (97) gives
\begin{eqnarray}
\lefteqn{-\frac{\Delta_{\rho}^{\prime\prime}}{2}-\frac{\Delta_{\mu}^{\prime\prime}}{2}+\frac{PP^{\prime\prime}}{\Delta_{\rho}}-\frac{2PP^{\prime}\Delta_{\rho}^{\prime}}{\Delta_{\rho}^{2}}-\frac{P^{2}\Delta_{\rho}^{\prime\prime}}{2\Delta_{\rho}}+\frac{P^{2}\Delta_{\rho}^{\prime^{2}}}{\Delta_{\rho}^{3}}+\frac{P^{\prime^{2}}}{\Delta_{\rho}}=}\hspace{8cm}\nonumber \\
& &2\Lambda A^{2}(\rho^{2}+a^{2}\mu^{2})\hspace{1cm}
\end{eqnarray}
Evidently (98) is separated to two parts ,one a function of $\mu$ only and the other a single-variable function of $\rho$ merely.So we can write
\begin{eqnarray}
& &\frac{\Delta_{\mu}^{\prime\prime}}{2}+2\Lambda a^{2}A^{2}\mu^{2}=C \\
& &-\frac{\Delta_{\rho}^{\prime\prime}}{2}+\frac{PP^{\prime\prime}}{\Delta_{\rho}}-\frac{2PP^{\prime}\Delta_{\rho}^{\prime}}{\Delta_{\rho}^{2}}-\frac{P^{2}\Delta_{\rho}^{\prime\prime}}{2\Delta_{\rho}}+\frac{P^{2}\Delta_{\rho}^{\prime^{2}}}{\Delta_{\rho}^{3}}+\frac{P^{\prime^{2}}}{\Delta_{\rho}}=2\Lambda A^{2}\rho^{2}+C\hspace{1cm}
\end{eqnarray}
where $C$ is a constant to be fixed.On the other hand (99) easily shows that $\Delta_{\mu}$ has a quartic form.Consistency with the boundary conditions and the generalized Kerr metric (2) implies the following form
\begin{equation}
\Delta_{\mu}=A^{2}(1-\mu^2)\left(1+\frac{\Lambda}{3}a^{2}\mu^{2}\right)
\end{equation}
This means that in (99), C is fixed by
\begin{equation}
C=A^{2}\left(\frac{\Lambda}{3}a^{2}-1\right)
\end{equation}
Using (102) and integrating (100) with respect to $\rho$ gives
\begin{equation}
\frac{\Delta_{\rho}^{\prime}}{2}-\frac{PP^{\prime}}{\Delta_{\rho}}+\frac{P^{2}\Delta_{\rho}^{\prime}}{2\Delta_{\rho}^{2}}=-\frac{2\Lambda}{3}A^{2}\rho^{3}-\left(\frac{\Lambda}{3}a^{2}-1\right)A^{2}\rho^{2}+D
\end{equation}
$D$ is another constant of integration that should be fixed. By another integration of (103) with respect to $\rho$ and fixing $D=-MA^2$ and the new constant of integration by $a^{2}A^{2}$ to satisfy the initial conditions we come to 
\begin{equation}
\Delta_{\rho}-\frac{P^2}{\Delta_{\rho}}=A^{2}\left[-\frac{\Lambda}{3}(\rho^{4}+a^{2}\rho^{2})+\rho^{2}-2M\rho+a^{2}\right]
\end{equation}
It can be noticed that the right-hand side of (104)has the same form of $\Delta_{\rho}$ for the generalized Kerr metric in (2).
Eq. (104) has two solutions
\begin{eqnarray}
\Delta_{\rho}&=&\frac{A^2}{2}\left\{\left[-\frac{\Lambda}{3}(\rho^{4}+a^{2}\rho^{2})+\rho^{2}-2M\rho+a^{2}\right]\right.\nonumber \\
& &\left.\pm\sqrt{\left[-\frac{\Lambda}{3}(\rho^{4}+a^{2}\rho^{2})+\rho^{2}-2M\rho+a^{2}\right]^2+\frac{4P^2}{A^4}}\right\}
\end{eqnarray}
The acceptable solution for the large distances would be the plus sign.
The field equations cannot fix $A$, but by imposing the condition that the variable $\mu$ should asymptotically have the same role as an azimuthal angle, we find
\begin{equation}
A=\frac{1}{1+\frac{\Lambda}{3}a^2}
\end{equation}
\section{Conclusion}
We may summarize our results in the form of line element as 
\begin{eqnarray}
ds^2&=&(\rho^{2}+a^{2}\cos^{2}\theta)\left\{\frac{d\theta^2}{1+\frac{\Lambda}{3}a^{2}\cos^{2}\theta}\right.\nonumber \\
& &\left.+\left.\left[d\rho-\frac{P(1+\frac{\Lambda}{3}a^{2})(dt-a\sin^{2}\theta d\phi)}{\rho^{2}+a^{2}\cos^{2}\theta}\right]^2\right/{\Delta_\rho}\right\} \nonumber \\
& &+\sin^{2}\theta\left(\frac{1+\frac{\Lambda}{3}a^{2}\cos^{2}\theta}{\rho^{2}+a^{2}\cos^{2}\theta}\right)\left[\frac{adt-(\rho^{2}+a^{2})d\phi}{1+\frac{\Lambda}{3}a^{2}}\right]^2\nonumber \\
& &-\frac{\Delta_{\rho}}{\rho^{2}+a^{2}\cos^{2}\theta}\left[\frac{dt-a\sin^{2}\theta d\phi}{1+\frac{\Lambda}{3}a^{2}}\right]^2
\end{eqnarray}
where $P=\sqrt{\frac{\Lambda}{3}}\rho^3$ and
\begin{eqnarray}
\Delta_{\rho}&=&\frac{1}{2}\left\{\left[-\frac{\Lambda}{3}(\rho^{4}+a^{2}\rho^{2})+\rho^{2}-2M\rho+a^{2}\right]\right.\nonumber \\
& &\left.+\sqrt{\left[-\frac{\Lambda}{3}(\rho^{4}+a^{2}\rho^{2})+\rho^{2}-2M\rho+a^{2}\right]^2+4P^{2}(1+\frac{\Lambda}{3}a^2)^4}\right\}\nonumber \\
& &
\end{eqnarray}
We have started by taking $\rho\equiv R(t)r$ with $\frac{\dot{R}}{R}=\sqrt{\frac{\Lambda}{3}}$, for which the equivalent differential form is
\begin{equation}
d\rho=\sqrt{\frac{\Lambda}{3}}\,\rho\, dt+R\,dr=\frac{P}{\rho^2}dt+R\,dr
\end{equation}
For coming to our final conclusion we take the following transformation instead of (109),
\begin{equation}
d\rho=\frac{P}{Z_{\rho}}\,dt+R\,dr
\end{equation}
where $P$ and $R$ are the same as what defined before and $Z_{\rho}$ is given by (89). Inserting (110) in (107) leads to the final form of the line element

\begin{eqnarray}
ds^2&=&(\rho^{2}+a^{2}\cos^{2}\theta)\left\{\frac{d\theta^2}{1+\frac{\Lambda}{3}a^{2}\cos^{2}\theta}\right.\nonumber \\
& &\left.+\left.\left[R(t)dr+\frac{P(1+\frac{\Lambda}{3}a^{2})}{(r^2+a^2)}dt-\frac{P(1+\frac{\Lambda}{3}a^{2})(dt-a\sin^{2}\theta d\phi)}{\rho^{2}+a^{2}\cos^{2}\theta}\right]^2\right/{\Delta_\rho}\right\} \nonumber \\
& &+\sin^{2}\theta\left(\frac{1+\frac{\Lambda}{3}a^{2}\cos^{2}\theta}{\rho^{2}+a^{2}\cos^{2}\theta}\right)\left[\frac{adt-(\rho^{2}+a^{2})d\phi}{1+\frac{\Lambda}{3}a^{2}}\right]^2\nonumber \\
& &-\frac{\Delta_{\rho}}{\rho^{2}+a^{2}\cos^{2}\theta}\left[\frac{dt-a\sin^{2}\theta d\phi}{1+\frac{\Lambda}{3}a^{2}}\right]^2
\end{eqnarray}
Evidently, from (111) we have 
\begin{equation}
g_{_{tt}}=-\frac{\Delta_{\rho}-a^2\sin^{2}\theta\left(1+\frac{\Lambda}{3}a^{2}\cos^{2}\theta\right)-P^{2}(1+\frac{\Lambda}{3}a^{2})^4\left[\frac{\rho^{2}+a^{2}\cos^{2}\theta}{(\rho^2+a^2)(1+\frac{\Lambda}{3}a^{2})}-1\right]^2/\Delta_{\rho}}{(\rho^{2}+a^{2}\cos^{2}\theta)(1+\frac{\Lambda}{3}a^{2})^2}
\end{equation}
It can be checked that (112) may be equal to zero for some values of $\rho$ very close to (6). In addition (111) gives 
\begin{equation}
g_{_{rr}}=\frac{(\rho^{2}+a^{2}\cos^{2}\theta)R^2}{\Delta_{\rho}}
\end{equation}
which is finite for the whole range of $\rho>0$. In order to find the event horizon we need to find $g^{rr}$. This can be done by using (110):
\begin{equation}
g^{rr}=\frac{1}{R^{^2}}\,g^{\rho\rho}+\frac{P^2}{R^{^2}Z_{\rho}^{2}}\,g^{tt}-2\frac{P}{R^{^2}Z_{\rho}}\,g^{\rho t}
\end{equation}
Inserting (16) in (114) gives 
\begin{equation}
g^{rr}=\frac{\Delta_{\rho}+\frac{P^{2}(1+\frac{\Lambda}{3}a^2)^4\sin^{2}\theta}{(\rho^2+a^2)(1+\frac{\Lambda}{3}a^2\cos^{2}\theta)}}{R^{^{2}}(\rho^{2}+a^{2}\cos^{2}\theta)}
\end{equation}
Since $\Delta_{\rho}>0$ for $\rho>0$, then $g^{rr}$ is finite in the whole range of $\rho>0$. This evidently shows the obtained metric is free from any event horizon. The intrinsic singularity remains the same as (3), i.e., $x^2+y^2=a^2$. Therefore the criterion of our result with respect to the Kerr and Kerr-de Sitter metrics is the absence of event horizon.
One may argue that the nature of the singularity at the event horizons of Kerr spacetime is of coordinate type and may be avoided by making a suitable transformation to a new coordinate system.Thus we are not faced by a serious problem then what is the significance of the new obtained results? The point is that there is no guarantee the new coordinate system being eligible to serve as the comoving frame.Our results are regular in a common coordinate system which its ability to serve as a comoving frame had been confirmed.

\end{document}